\documentclass[a4paper,
twocolumn,
11pt,
unpublished]{quantumarticle}

\pdfoutput=1
\usepackage[utf8]{inputenc}
\usepackage[english]{babel}
\usepackage[T1]{fontenc}
\usepackage{amsmath, amsfonts, amssymb, amsthm}
\usepackage{bm}
\usepackage{hyperref}
\usepackage{tikz}
\usepackage{enumitem}
\newcommand{\subscript}[2]{$\mathrm{#1} _ #2$}
\usepackage{dsfont}
\usepackage[numbers, sort&compress]{natbib}

\usepackage[acronym]{glossaries}
\glsdisablehyper

\definecolor{iqmblue}{RGB}{117,157,235}
\definecolor{iqmgreen}{RGB}{95, 221, 151}

\newcommand*{\f}{\frac}
\newcommand*{\mc}{\mathcal}
\newcommand*{\dg}{\dagger}

\newcommand*{\mbb}{\mathds}

\newcommand*{\gatefid}{\mathsf{F}}
\newcommand*{\unitarity}{\mathsf{u}}
\newcommand*{\entfid}{\mathrm{F}_\mathsf{ent}}

\newcommand*{\twist}{\bm{\circlearrowleft}}
\newcommand*{\spark}{IQM Spark\textsuperscript{\tiny{TM}}}

\DeclareMathOperator{\tr}{tr}
\DeclareMathOperator*{\Motimes}{\text{\raisebox{0.25ex}{\scalebox{0.6}{$\bigotimes$}}}}
\DeclareMathOperator*{\Moplus}{\text{\raisebox{0.25ex}{\scalebox{0.6}{$\bigoplus$}}}}

\DeclareMathOperator*{\expect}{\mathop{\mbb{E}}}

\newacronym{rb}{RB}{Randomized Benchmarking}
\newacronym{spam}{SPAM}{State Preparation and Measurement}
\newacronym{asf}{ASF}{Average Sequence Fidelity}
\newacronym{udd}{UDD}{Universal Dynamical Decoupling}
\newacronym{cp}{CP}{Completely Positive}
\newacronym{tp}{TP}{Trace Preserving}
\newacronym{cptp}{CPTP}{Completely Positive Trace Preserving}
\newacronym{cptni}{CPTNI}{Completely Positive Trace Non-Increasing}
\newacronym{ptm}{PTM}{Pauli Transfer Matrix}
\newacronym{povm}{POVM}{Positive Operator Valued Measurement}
\newacronym{rc}{RC}{Randomized Compiling}
\newacronym{mrb}{MRB}{Mirror Randomized Benchmarking}
\newacronym{birb}{BiRB}{Binary Randomized Benchmarking}
\newacronym{drb}{DRB}{Direct Randomized Benchmarking}
\newacronym{urb}{URB}{Unitarity Randomized Benchmarking}
\newacronym{rm}{RM}{Randomized Measurements}
\newacronym{moms}{MoMs}{Median of Means}

\theoremstyle{definition}
\newtheorem{result}{Result}

\begin{document}
\title{Estimating the coherence of noise in mid-scale quantum systems}

\author{Pedro Figueroa-Romero}
\email{pedro.romero@meetiqm.com}
\affiliation{IQM Quantum Computers, Georg-Brauchle-Ring 23-25, 80992 Munich, Germany}

\author{Miha Papi\v{c}}
\affiliation{IQM Quantum Computers, Georg-Brauchle-Ring 23-25, 80992 Munich, Germany}
\affiliation{%
 Department of Physics and Arnold Sommerfeld Center for Theoretical Physics,\\
Ludwig-Maximilians-Universit\"at M\"unchen, Theresienstr. 37, 80333 Munich, Germany
}%

\author{Adrian Auer}
\affiliation{IQM Quantum Computers, Georg-Brauchle-Ring 23-25, 80992 Munich, Germany}

\author{Inés de Vega}
\affiliation{IQM Quantum Computers, Georg-Brauchle-Ring 23-25, 80992 Munich, Germany}
\affiliation{%
 Department of Physics and Arnold Sommerfeld Center for Theoretical Physics,\\
Ludwig-Maximilians-Universit\"at M\"unchen, Theresienstr. 37, 80333 Munich, Germany
}%

%\date{\today}

\begin{abstract}
While the power of quantum computers is commonly acknowledged to rise exponentially, it is often overlooked that the complexity of quantum noise mechanisms generally grows much faster. In particular, quantifying whether the instructions on a quantum processor are close to being unitary has important consequences concerning error rates, e.g., for the confidence in their estimation, the ability to mitigate them efficiently, or their relation to fault-tolerance thresholds in error correction. However, the complexity of estimating the coherence, or unitarity, of noise generally scales exponentially in system size. Here, we obtain an upper bound on the average unitarity of Pauli noise and develop a protocol allowing us to estimate the average unitarity of operations in a digital quantum device efficiently and feasibly for mid-size quantum systems. We demonstrate our results through both experimental execution on \spark, a 5-qubit superconducting quantum computer, and in simulation with up to 10 qubits, discussing the prospects for extending our technique to arbitrary scales.
\end{abstract}

%\tableofcontents

The main obstacle for quantum computing undeniably remains the presence of noise, causing a multitude of errors that limit the usefulness of currently available quantum processors. That is to say, not all quantum noise is made equal, and ironically, figures of merit that are efficient to access in an experiment --such as the fidelity of a quantum operation-- are often by definition insensitive to many of the details of such noise. It is thus imperative to have a plethora of protocols with certain practical desiderata that can probe different aspects of noise, and that can then, in turn, provide actionable information to existing methods to suppress, mitigate, or fully correct potential errors.

Quantum noise can be described as an undesirable quantum transformation mapping the ideal noiseless output state of a quantum computation into some other valid, albeit unexpected, output quantum state. Focusing on the regime where such noise is \emph{effectively} Markovian, i.e., it only depends on the input state and no other external context variables, a comprehensive classification of all possible errors can be made~\cite{PRXQuantum.3.020335}, but in particular it is extremely important to be able to distinguish between \emph{coherent} and \emph{incoherent} error contributions.

Coherent noise can be described as a deterministic, undesired unitary transformation, e.g., it can arise due to imperfect control or calibration of quantum gates, and also due to certain types of crosstalk~\cite{PRXQuantum.2.040338}. Incoherent or stochastic noise, on the other hand, is described by non-unitary transformations and arises purely statistically, e.g., a bit-flip occurring with a certain non-zero probability, or so-called \emph{depolarizing noise}, whereby a state gets maximally mixed with a certain probability but remains intact with the complementary probability. Generally, Markovian quantum noise will be described by a quantum operation that contains both coherent and stochastic noise contributions. The reasons why distinguishing between these is important, include:
\begin{enumerate}[nolistsep, label=\emph{\roman*})]
    \item Coherent errors can, for example, be effectively suppressed by good experimental quantum control, while incoherent errors in isolation can be efficiently mitigated~\cite{Temme_PEC_2017, McDonough_2022, Gonzales_2023, PhysRevResearch.5.033193, vandenBerg_PEC_2023} and fully corrected~\cite{knill2005, eastin2007error}.
    \item  Coherent errors can accumulate in a much more detrimental way and provide error-rate estimates, such as average gate-fidelity, that can differ by orders of magnitude from worst-case estimates (i.e., fault-tolerance thresholds)~\cite{Wallman_2014, Sanders_2015, Wallman_2015, Kueng_2016, Hashim_2023}.
    \item The interpretation of average gate fidelity as a reliable and meaningful error rate depends heavily on the noise having a low coherent contribution~\cite{Wallman_2015, PhysRevA.94.052325, unitarity_helsen_2019, Hashim_2021}.
\end{enumerate}

\begin{figure}[t!]
    \centering
    \includegraphics[width=0.5\textwidth]{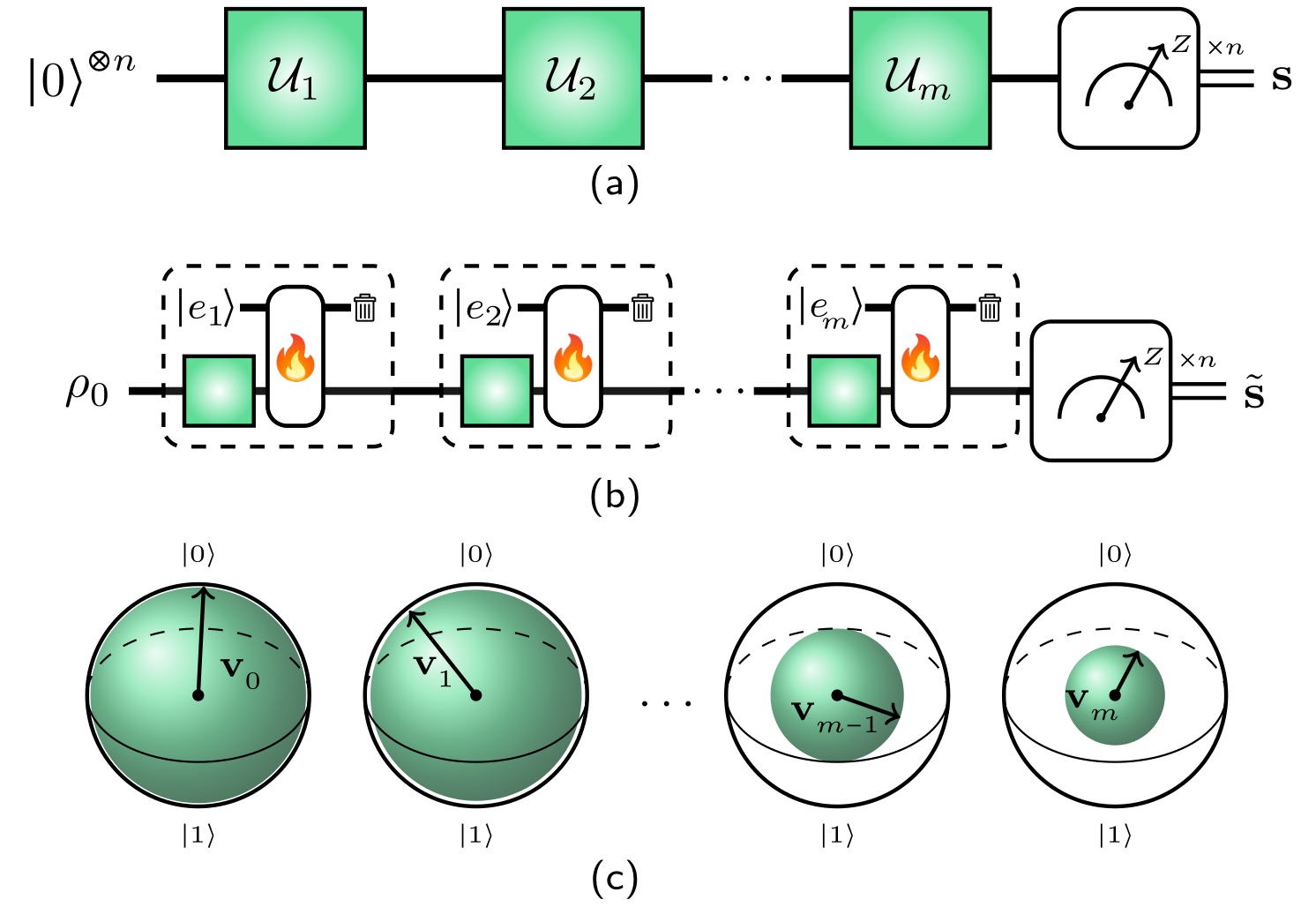}
    \caption{\textbf{Coherence of noise and loss of purity}: (a) An ideal quantum computation on $n$ qubits can be described by an initial state $|0\rangle^{\otimes{n}}$, a sequence of unitary operations $\mc{U}_i$, and a computational basis measurement with output $\mathbf{s}\in\{0,1\}^n$. More realistically, in (b), the initial state can be statistically mixed and correlated among qubits, and operations are subject to noise due to interactions with external degrees of freedom $|e_i\rangle$ that further reduce the state purity. In (c), the impurity of each state $\rho_i:=\f{1}{2^n}\left(\mbb1+\mathbf{v}_i\cdot\mathbf{P}\right)$ after the $i$\textsuperscript{th} operation, where $\mathbf{P}$ is a vector of all $n$-qubit Pauli matrices and $\cdot$ an inner product, can be depicted as a shrinkage of the $2^{n-1}$-dimensional Bloch vector, $|\mathbf{v}_0|\geq|\mathbf{v}_1|\geq\ldots\geq|\mathbf{v}_m|$; such shrinking can be associated on average to the \emph{unitarity} or \emph{coherence} amount of the noise, which is a resource-intensive quantity to characterize experimentally as $n$ increases.}
    \label{fig: teaser}
\end{figure}

The average coherence of a noisy quantum gate can be understood as a measurement of the rate at which noise shrinks the $n$-dimensional Bloch ball, as depicted in Fig.~\ref{fig: teaser}, and can be measured by its \emph{unitarity}, which in turn can be estimated operationally by \gls{urb}~\cite{Wallman_2015}. Despite being efficient and robust to \gls{spam} errors, \gls{urb} is not scalable in system size, as it requires estimating an exponential amount of expectation values within a \gls{rb}-like protocol. The reason for this is not \gls{urb} \emph{per se}, but rather that fundamentally, the unitarity is a second-order functional of the noise, akin to the case of the so-called \emph{purity} for a quantum state.

In general, the complexity of characterizing quantum noise comprehensively typically scales exponentially with system size~\cite{certif_and_bench_2020, PRXQuantum.2.010201}. Even the most efficient of protocols estimating a single figure of merit, such as those within the family of \gls{rb}~\cite{helsen_general}, fail to do so at scale. However, with the quantum industry inevitably moving towards larger systems, there is a pressing need to find ways to generalize such tools to larger systems.

Recently, scalable \gls{rb}-based techniques such as \gls{mrb}~\cite{mrb_prl2022, hines2022demonstrating} and \gls{birb}~\cite{hines2023fully}, among others~\cite{mckay2023benchmarking, hines2023scalable}, have been developed, which can estimate the average fidelity of sets of quantum operations in a scalable, efficient and \gls{spam}-robust way, for a large class of gate sets. On the other hand, \gls{rm} techniques, and so-called classical shadows, have enabled the efficient estimation of properties of many-body quantum systems~\cite{shadows_2020, rmtoolbox_2022} (including state fidelity and purity), linear properties of gate sets~\cite{helsen2021estimating}, and characterization of processes with memory~\cite{PhysRevLett.130.160401}.

We harness the results of~\cite{vanEnk_2012, Renyi_2018, purity_science_2019, purity_pra_2019}, together with a scalable \gls{rb}-inspired protocol, to develop a standalone technique allowing to simultaneously estimate the average fidelity and average unitarity of layers of operations in mid-scale quantum systems. Moreover, we first identify an interval for the unitarity of so-called stochastic Pauli noise in terms of average fidelity, enabling certification of whether average noise has a low coherent contribution and establishing meaningful error budgets via the average unitarity.

In \S~\ref{sec: background} we introduce the main technical background, while in \S~\ref{sec: Pauli noise unitarity} we present a new upper bound on the unitarity of Pauli noise. In \S~\ref{sec: rms} we summarize existing results regarding \gls{rm}s, before presenting our protocol in \S~\ref{sec: main}, which in \S~\ref{sec: demonstration} we demonstrate with experimental execution on \spark~\cite{spark}. In the remaining \S~\ref{sec: scaling bottlenecks} we discuss the bottlenecks for a scalable estimation of average coherence of noise, and finally, we draw some conclusions and prospects to overcome mid-scalability in \S~\ref{sec: conclusions}.

\section{Average coherence of noise and its relation with purity and fidelity}\label{sec: background}
It is well known that no physical system can be perfectly isolated. In particular, the loss of coherence of quantum states is one of the fundamental practical problems facing any quantum technology. In general, the dissipation of information to an external environment results in an increasingly mixed classical probability distribution over a set of possible quantum states. Such mixedness or uncertainty can be quantified by the purity of the respective density matrix $\rho$, given by $\tr(\rho^2)$. The purity takes extremal values of 1 for a pure state, and $2^{-n}$ for a maximally mixed state on $n$-qubits, and it can also be written as the Schatten 2-norm $\|\rho\|_2^2$, where $\|X\|_2=[\tr(XX^\dg)]^{1/2}$.

Logical operations or gates $\mc{G}$ in a quantum computer are described by unitary operators, which by definition preserve the purity of quantum states, i.e., for $\mc{G}(\rho_{\mathsf{in}})=\rho_\mathsf{out}$, we have $\tr(\rho_\mathsf{in}^2)=\tr(\rho_\mathsf{out}^2)$. The corresponding, real noisy operation, $\mc{G}_\mathsf{noisy}$, however, is described by a more general \gls{cp} map, such that for $\mc{G}_\mathsf{noisy}(\rho_\mathsf{in})=\rho_\mathsf{out}$, we have $\tr(\rho_\mathsf{out}^2)\leq\tr(\rho_\mathsf{in}^2)$. The reduction in purity of states when acted on by a noisy channel is related to the \emph{unitarity} of such channel, a measure of ``\emph{how far}'' $\mc{G}_\mathsf{noisy}$ is from being unitary. Because locally we can always relate a noisy channel with an ideal gate by $\mc{G}_\mathsf{noisy}:=\mc{E}\circ\mc{G}$~(\cite{PRXQuantum.3.020335}) for some \gls{cp} map $\mc{E}$, the unitarity of the noisy gate equivalently measures how far $\mc{E}$ is from being unitary.

The average unitarity of a quantum channel can be defined as in~\cite{Wallman_2015}, by
\begin{align}
    \overline{\unitarity}(\mc{E}) &:= \f{2^n}{2^n-1}\expect_{\psi\sim\mathsf{Haar}}\|\mc{E}^\prime(\psi)\|_2^2,
    \label{eq: def uniform avg unitarity}
\end{align}
where $\mc{E}^\prime(\cdot) := \mc{E}(\,\cdot-\mbb1/2^n)$, with $\mbb1$ being an $n$-qubit identity operator, and the normalization $2^n/(2^n-1)$ ensuring that $0\leq\overline{\unitarity}(\mc{E})\leq1$, and with $\overline{\unitarity}(\mc{E})=1$ if and only if $\mc{E}$ ---and hence the noisy operation $\mc{G}_\mathsf{noisy}$--- is unitary. Since the purity of a noisy output state may be written in the form $\tr(\rho_\mathsf{out}^2)=\|\mc{E}(\rho_{\mathsf{in}})\|_2^2$, the definition in Eq.~\eqref{eq: def uniform avg unitarity} implies that the average purity of a noisy state is, in general, a combination of average unitarity and trace-decrease and/or non-unital terms~\footnote{ See, e.g., Eq.~\eqref{eq: avg purity m=1 general spam} for an explicit expression.}.

The unitarity only quantifies \emph{how unitary} $\mc{G}_\mathsf{noisy}$ is, but not whether it is the target unitary, i.e., it does not measure whether $\mc{G}_\mathsf{noisy}$ is far from the ideal $\mc{G}$. A figure of merit aiming to quantify this distinction is the average gate fidelity,
\begin{align}
    \overline{\mathsf{F}}(\mc{G}_\mathsf{noisy},\mc{G}) &:= \expect_{\psi\sim\mathsf{Haar}}\tr[\mc{G}_\mathsf{noisy}(\psi)\mc{G}(\psi)] \nonumber\\
    &= \expect_{\psi\sim\mathsf{Haar}}\langle\psi|\mc{E}(\psi)|\psi\rangle := \overline{\mathsf{F}}(\mc{E}),
\end{align}
where implicitly we denote $\overline{\mathsf{F}}(\mc{E}):=\overline{\mathsf{F}}(\mc{E},\mc{I})$, the average gate fidelity of $\mc{E}$ with respect to the identity $\mc{I}$ (and in a slight abuse of notation, $\psi=|\psi\rangle\!\langle\psi|$). The average gate fidelity satisfies $1/(2^n+1)\leq\overline{\mathsf{F}}(\mc{E})\leq1$, with $\overline{\mathsf{F}}(\mc{E})=1$ if and only if $\mc{E}=\mc{I}$ (equivalently if and only if $\mc{G}_\mathsf{noisy}=\mc{G}$). With respect of $\overline{\gatefid}$, in~\cite{Wallman_2015} it is shown that the average unitarity satisfies $\overline{\unitarity}(\mc{E}) \geq \overline{f}(\mc{E})^2$, where
\begin{equation}
    \overline{f}(\mc{E}) = \f{2^n\overline{\gatefid}(\mc{E})-1}{2^n-1},
    \label{eq: average polarization}
\end{equation}
is the so-called average polarization of $\mc{E}$, and with saturation occurring for a depolarizing (purely incoherent) channel.

The unitarity can also be written in terms of fidelity as $\overline{\unitarity}(\mc{E}) = (2^n/(2^n-1))\overline{\mathsf{F}}(\mc{E}^{\prime\,\dg}\mc{E}^\prime)$, where $\mc{E}^{\prime\,\dg}$ is the adjoint map of $\mc{E}^\prime$\footnote{ The adjoint map $\Phi^\dg$ of a \gls{cp} map $\Phi$ is defined by $\tr[A\Phi(B)]=\tr[\Phi^\dg(A)B]$, or equivalently by conjugating the Kraus operators of $\Phi$.}. Thus, the average unitarity can also be understood as a type of second-order function of the fidelity of $\mc{E}^\prime$, quantifying on average how distinguishable is the composition $\mc{E}^{\prime\,\dg}\mc{E}^\prime$ from the identity. Under the assumptions that $\mc{E}$ approximately models the average noise of a gate set $\{\mc{G}_i\}$ in a temporally uncorrelated (Markovian), time-independent and gate-independent way, $\overline{\mathsf{F}}(\mc{E})$ can be estimated through \gls{rb}, and the unitarity $\overline{\unitarity}(\mc{E})$ can be estimated through \gls{urb}~\cite{Wallman_2015, unitarity_helsen_2019}. Neither technique is scalable in the number of qubits, however, recently the \gls{mrb} and \gls{birb} variants were developed in~\cite{mrb_prl2022, hines2022demonstrating, hines2023fully} allowing efficient and scalable estimation of at least $\overline{\mathsf{F}}(\mc{E})$ for a large number of qubits.

\section{Pauli noise unitarity}\label{sec: Pauli noise unitarity}
Incoherent noise is \gls{tp} and unital, and a distinction from coherent noise is that it generates no net rotation of the state space~\cite{PRXQuantum.3.020335}. Pauli noise is a special case of incoherent noise described by a channel $\rho\mapsto\sum_{P\in\mathsf{Pauli}_n}\alpha_PP\rho{P}$, where $\alpha_P$ is the Pauli error probability associated to a $n$-qubit Pauli operator $P$. Its relevance stems not only from that of the Pauli basis in quantum information theory and experiment but also from its extensive application in quantum error characterization~\cite{PhysRevLett.106.230501, PhysRevLett.107.253602, Flammia_2020, Harper_2020, PRXQuantum.2.010322, Chen_2023, berg2023techniques}, quantum error mitigation~\cite{Temme_PEC_2017, McDonough_2022, Gonzales_2023, PhysRevResearch.5.033193, vandenBerg_PEC_2023}, quantum error correction~\cite{eastin2007error, PhysRevLett.121.190501, Wagner2022paulichannelscanbe} and in its connection with fault tolerance thresholds~\cite{PhysRevA.91.022335, PhysRevLett.120.050505, Sanders_2015, Kueng_2016,Hashim_2023}. Any quantum channel $\mc{E}$ can be averaged to a corresponding Pauli channel $\mc{E}^{\twist}$ by means of its Pauli twirl, $\mc{E}\mapsto\mc{E}^{\twist}(\cdot):=\sum_{P\in\mathsf{Pauli}_n}{4}^{-n}P\mc{E}(P\cdot{P})P$, i.e., reduced to a Pauli channel with the same Pauli error probabilities as $\mc{E}$.

\begin{result}[\emph{Fidelity bound for Pauli unitarity}]\label{result: Pauli unitarity} The unitarity of the \emph{Pauli-twirled} channel $\mc{E}^{\twist}$ of $\mc{E}$ is bounded by
\begin{equation}
    \overline{f}(\mc{E})^2 \leq \overline{\unitarity}(\mc{E}^{\twist}) \leq \f{4^n-2}{(2^n-1)^2}\,\overline{r}(\mc{E})^2+\overline{f}(\mc{E})^2,
    \label{eq: main Pauli unitarity bounds}
\end{equation}
where $\overline{f}$ is average polarization defined by Eq.~\eqref{eq: average polarization}, and $\overline{r}:=1-\overline{\gatefid}$ is average infidelity.
\end{result}

\begin{figure}[t!]
    \centering
    \includegraphics[width=0.5\textwidth]{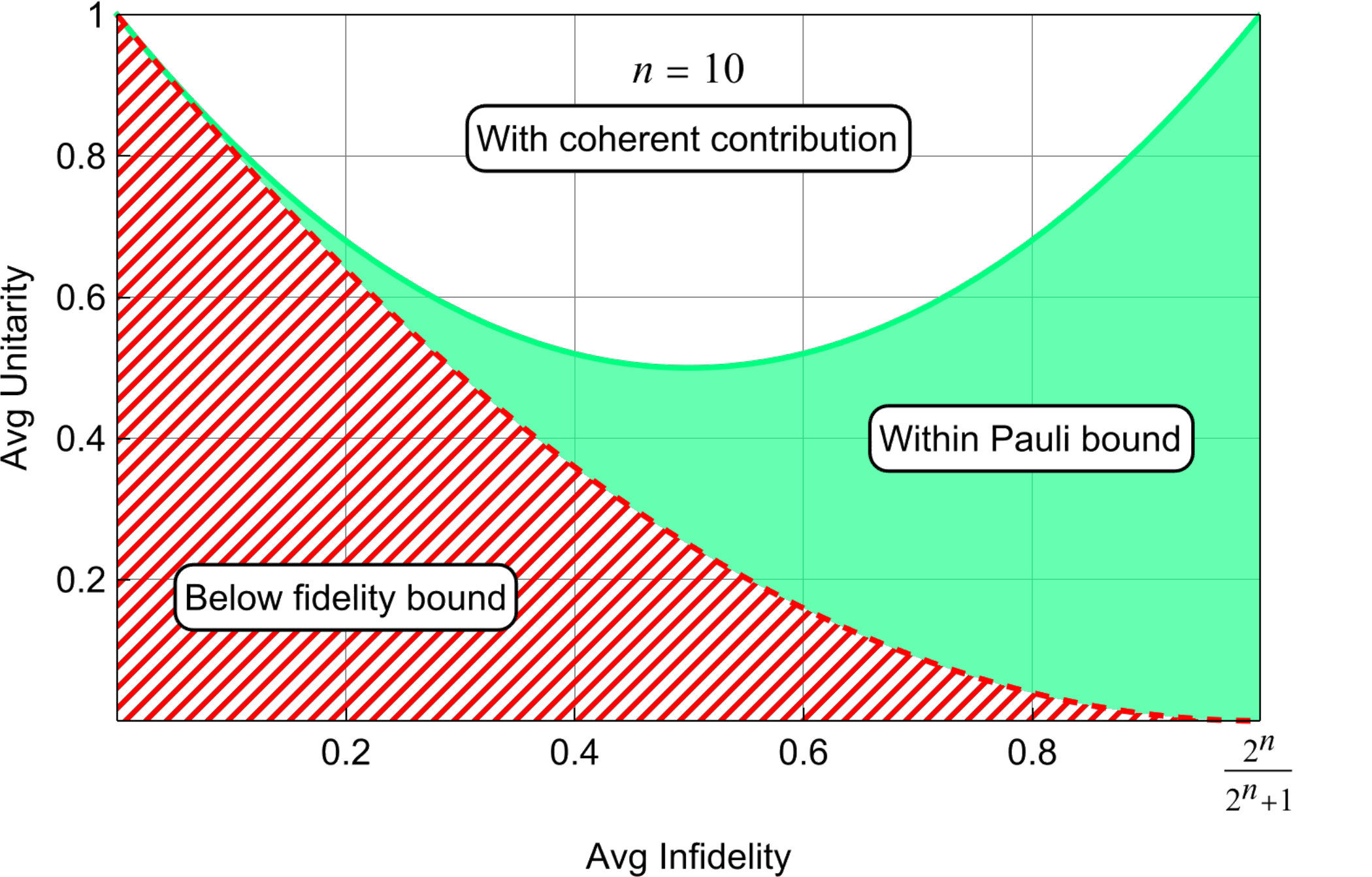}
    \caption{\textbf{Pauli unitarity region with respect to infidelity (case $n=10$ qubits).} Regions for all possible pairs of average unitarity and fidelity: the dashed red region denotes all impossible combinations for Markovian \gls{cp} noise, the green shaded region denotes values within Ineq.~\eqref{eq: main Pauli unitarity bounds} for the average unitarity of Pauli noise, and the white region denotes all values outside such bound, guaranteed to contain coherent error contributions.}
    \label{fig: pauli purity bound}
\end{figure}

The proof is detailed in in Appendix~\ref{appendix: Pauli stuff}. Given an average layer (in)fidelity, the bound in Ineq.~\eqref{eq: main Pauli unitarity bounds} allows us to estimate whether average noise is close to Pauli through the average unitarity. Furthermore, the Pauli-twirled channel $\mc{E}^{\twist}$ of a given \emph{noise} channel $\mc{E}$ can be efficiently approximated operationally by \gls{rc}~\cite{PhysRevA.94.052325, Hashim_2021}, so Ineq.~\eqref{eq: main Pauli unitarity bounds} enables to certify \gls{rc} through the average unitarity. Generally, for small average infidelity, Ineq.~\eqref{eq: main Pauli unitarity bounds} will be tight, and the average unitarity will need to approach the square of the average polarization, $\overline{f}(\mc{E})^2$, for the average noise to be strictly Pauli. This can be seen more clearly for a fixed number of qubits $n$, as in Fig.~\ref{fig: pauli purity bound}.

The upper bound in Ineq.~\eqref{eq: main Pauli unitarity bounds} overestimates the average unitarity of Pauli noise by a proportion of cross-products of non-identity Pauli error rates, $(\sum\alpha_P)^2-\sum\alpha_P^2$, and further assumes \gls{tp} noise, so while a unitarity outside the Pauli bounds guarantees that average noise will contain coherent error contributions, a unitarity value within the Pauli interval~\eqref{eq: main Pauli unitarity bounds} solely points to the likelihood of the average noise being Pauli.

\section{Purity and fidelity via randomized measurements}\label{sec: rms}
The main roadblock for a scalable \gls{urb} stems from purity being a quadratic function of a quantum state, in general requiring knowledge of all its components, e.g., as described in~\cite{Wallman_2015} requiring state tomography within a \gls{rb} protocol. The efficient estimation of purity, however, was one of the first problems giving rise to the \gls{rm} and shadow tomography techniques~\cite{vanEnk_2012, Renyi_2018, rmtoolbox_2022}. While for the particular case of purity, the number of required measurements for a given accuracy still scales exponentially, the exponent for \gls{rm}s is significantly smaller than for full tomography~\cite{ rmtoolbox_2022}.

We will focus on the result of~\cite{purity_science_2019, purity_pra_2019},
\begin{equation}
    \tr(\rho^2\,) = 2^{n}\sum_{\mathbf{s},\mathbf{{s}^\prime}} (-2)^{-h(\mathbf{s},\mathbf{s}^\prime)}\expect_{\mathsf{U}_i\sim\mathsf{Haar}}\mathsf{P}_\mathsf{U}^{(\mathbf{s})}\mathsf{P}_\mathsf{U}^{(\mathbf{s}^\prime)},
    \label{eq: purity random cross-correlations}
\end{equation}
where here $\mathsf{U}=\Motimes_{i=1}^{n}\mathsf{U}_i$ is a random unitary, with $\expect_{\mathsf{U}_i\sim\mathsf{Haar}}$ denoting averaging with the Haar measure over the each $\mathsf{U}_i$, $\mathbf{s}$ and $\mathbf{s}^\prime$ are $n$-bit strings, $h(\mathbf{s},\mathbf{s}^\prime)$ is the Hamming distance (number of distinct bits) between them, and the
\begin{equation}
\mathsf{P}_\mathsf{U}^{(\mathbf{s})} := \langle\mathbf{s}|\mathsf{U}^\dg\rho\,\mathsf{U}|\mathbf{s}\rangle,
    \label{eq: probability purity orig}
\end{equation}
are probabilities of observing the $n$-bit string $\mathbf{s}$ upon randomizing with $\mathsf{U}$, and which are estimated in experiment. Since Eq.~\eqref{eq: purity random cross-correlations} involves two copies of $\mathsf{U}$ and $\mathsf{U}^\dg$, it suffices that the local unitaries $\mathsf{U}_i$ belong to a unitary 2-design~\cite{PhysRevA.80.012304}, such as the uniformly-distributed Clifford group.

By a \gls{rm}, here we will mean a random unitary operation $\mathsf{U}$, followed by a projective measurement in the computational basis, $\mathsf{U}|\mathbf{s}\rangle$~\footnote{ More strictly, this corresponds to a \gls{rm} \emph{element}, and the concept of a \gls{rm} can be formalized with Random \gls{povm}s, as in~\cite{Heinosaari_2020}}. In Eq.~\eqref{eq: probability purity orig}, this can equivalently be read as randomizing the state $\rho$ with $\mathsf{V}=\mathsf{U}^\dg$, and then performing a computational basis measurement.

Purity can alternatively be estimated through the so-called \emph{shadow} of the state, constructed via \gls{rm}s. That is, one can equivalently construct proxy states $\hat{\rho}_\mathsf{s}$ through the probabilities $\mathsf{P}_\mathsf{U}(\mathsf{s})$ and compute the purity through products $\tr(\hat{\rho}_\mathsf{s}\hat{\rho}_{\mathsf{s}^\prime})$~\cite{shadows_2020}. This is an equivalent approach with similar performance guarantees~\cite{rmtoolbox_2022}; here we employ Eq.~\eqref{eq: purity random cross-correlations} because it can be used within a \gls{rb}-like protocol straightforwardly.

Operationally, the purity is truly important once in light of an associated fidelity~\cite{Di_Franco_2013}: notice that the probabilities $\mathsf{P}_\mathsf{U}^{(\mathbf{s})}$ can be \emph{recycled} in a straightforward way to estimate the fidelity of $\rho$ with respect to an ideal pure state $|\psi\rangle$, through a slight modification to Eq.~\eqref{eq: purity random cross-correlations}, as
\begin{equation}
    \langle\psi|\rho|\psi\rangle = 2^{n}\sum_{\mathbf{s},\mathbf{{s}^\prime}} (-2)^{-h(\mathbf{s},\mathbf{s}^\prime)}\expect_{\mathsf{U}_i\sim\mathsf{Haar}}\!\!\mathsf{P}_\mathsf{U}^{(\mathbf{s})}\mathsf{Q}_\mathsf{U}^{(\mathbf{s}^\prime)},
\end{equation}
where here now $\mathsf{Q}_\mathsf{U}^{(\mathbf{s})}=|\langle\psi|\mathsf{U}|\mathbf{s}\rangle|^2$ is the probability of observing the $n$-bit string $\mathbf{s}$ upon randomizing with $\mathsf{U}=\Motimes_i\mathsf{U}_i$ and measuring $|\psi\rangle$. This comes at the expense of estimating the ideal probabilities $\mathsf{Q}_\mathsf{U}^{(\mathbf{s})}$, but this can be done classically in an efficient way since the $\mathsf{U}_i$ can be replaced by single-qubit Clifford unitaries.

\section{Average unitarity through randomized measurement correlations}\label{sec: main}
While we do not require the same setup as in \gls{mrb} or \gls{birb}, we mainly follow the framework and notation of~\cite{hines2022demonstrating}. We will, in particular, only consider single-qubit and two-qubit gate sets, $\mbb{G}_1$ and $\mbb{G}_2$, distributed according to probability distributions $\Omega_1$ and $\Omega_2$, respectively. We will refer to instructions with these gate sets on $n$-qubits, simply as \emph{layers}, and denote the corresponding set of layers by $\mbb{L}(\mbb{G})$, where $\mbb{G}=\mbb{G}_1\cup\mbb{G}_2$. In practice, a rule on how to apply the sampled gates must also be specified, depending e.g., on the topology, connectivity, and the desired density of the gates.

We construct $n$-qubit random circuits, of layer circuit circuit depth $m$, of the form
\begin{equation}
    \mathsf{C}_m=\mathsf{L}_m\mathsf{L}_{m-1}\cdots\mathsf{L}_1,
    \label{eq: def omega-distributed circuit}
\end{equation}
where the right-hand side is read (or acts) from right to left, and where each $\mathsf{L}_i=\mathsf{L}_i^{(2)}\mathsf{L}_i^{(1)}$ is a composite layer, made up of subsequent applications of a layer $\mathsf{L}_i^{(1)}\in\mbb{L}(\mbb{G}_1)$, of only parallel single-qubit gates, and a layer $\mathsf{L}_i^{(2)}\in\mbb{L}(\mbb{G}_2)$, of only parallel two-qubit gates. We will say that all $\mathsf{L}_i\in\mbb{L}(\mbb{G})$ are random layers distributed according to $\Omega$, which is such that $\Omega(\mathsf{L}_i)=\Omega_1(\mathsf{L}_i^{(1)})\Omega_2(\mathsf{L}_i^{(2)})$, and we refer to $\mathsf{C}_m$ in Eq.~\eqref{eq: def omega-distributed circuit} as a \emph{$\Omega$-distributed random circuit of depth $m$}.

\begin{figure}[t!]
    \centering
    \includegraphics[width=0.495\textwidth]{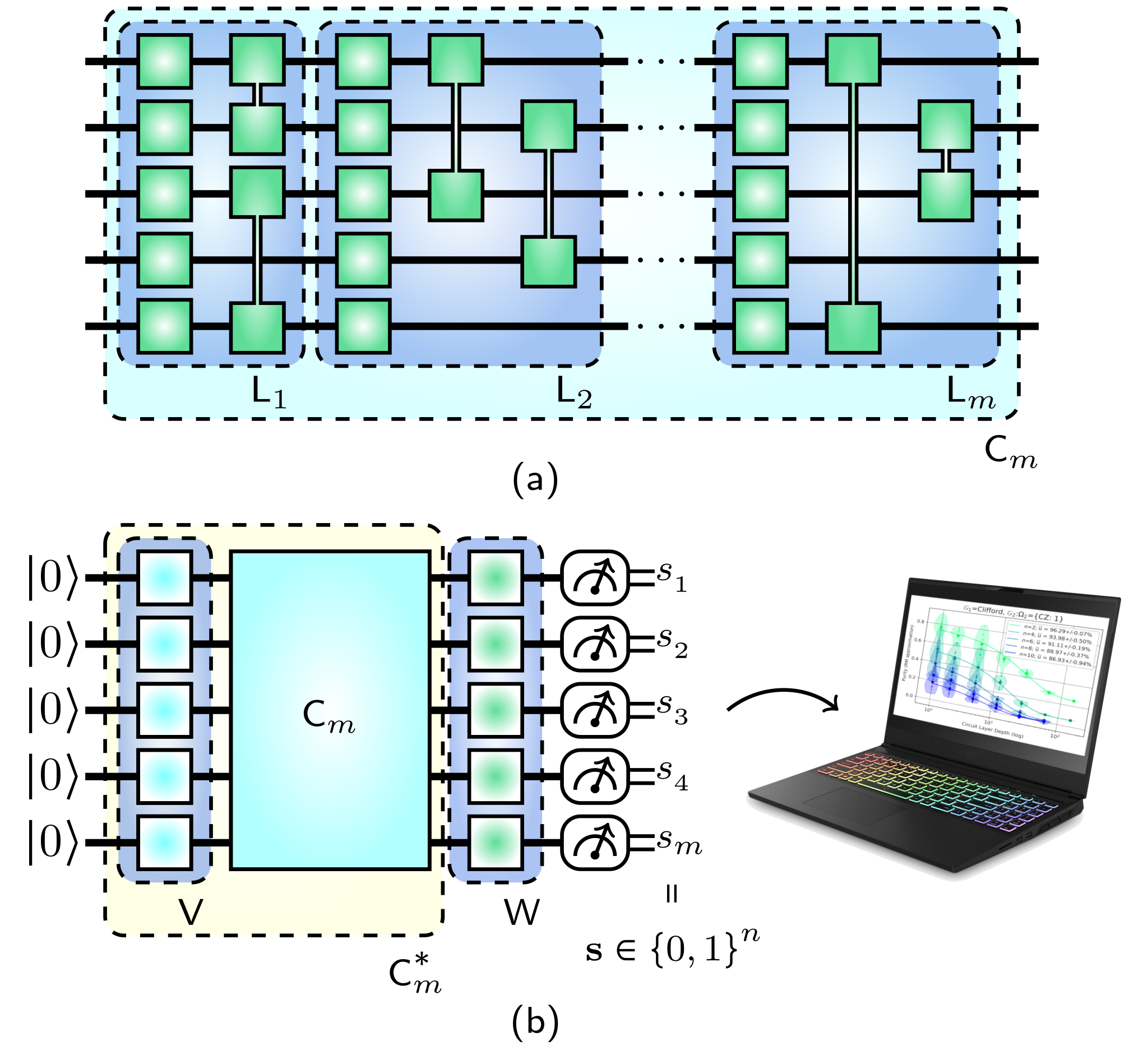}
    \caption{\textbf{$\Omega$-distributed circuits and RMs}: In $\mathsf{(a)}$, a sketch of a sample of a ($n=5$ qubit) $\Omega$-distributed circuit $\mathsf{C}_m$, as a sequence of random layers $\mathsf{L}_1,\mathsf{L}_2,\ldots,\mathsf{L}_m\sim\Omega$ made up of only parallel (according to device topology) layers of single- and two-qubit gates, $\mbb{G}_1$ and $\mbb{G}_2$, respectively, sampled according to their corresponding distributions $\Omega_1$ and $\Omega_2$. In $\mathsf{(b)}$, a layer of random single-qubit Clifford gates $\mathsf{V}$ is preppended to the circuit $\mathsf{C}_m$, defining a circuit $\mathsf{C}_m^*$, and the final measurement is randomized with a different layer of random single-qubit Clifford gates $\mathsf{W}$, generating a $n$-bit string $\mathbf{s}$; a set of such measurements can be processed according to Protocol~\ref{sec: protocol} to estimate the average unitarity of the corresponding $\Omega$-distributed circuits.}
    \label{fig: Random Circuits and RMS}
\end{figure}

We use the term \emph{depth} to mean the number of layers entering the given $\Omega$-distributed circuit, \emph{before} transpilation to a given basis gate set. We illustrate these concepts in Fig.~\ref{fig: Random Circuits and RMS}. We also point out that while the choice of the pair $(\mbb{G},\Omega)$ is a priori arbitrary, here we will only consider $\mbb{G}_1:=\mathsf{Clif}_1$ being the Clifford group (or effectively any other efficient single-qubit unitary 2-design) and $\Omega_1$ a uniform distribution.

Finally, below we distinguish estimators of expectation values of a random variable by a circumflex, i.e., $\hat{\mathsf{X}}$ represents the estimator of the expectation of some random variable $\mathsf{X}$.

\subsection{The protocol}\label{sec: protocol}
With this setup, the protocol consists of:
\begin{enumerate}[wide, leftmargin=*, labelwidth=!, labelindent=0pt]
    \item\label{protocol: preparation} \emph{Preparation}: Generate $N_{\mathsf{C}_m^*}$ samples of $n$-qubit $\Omega$-distributed circuits $\mathsf{C}_m$ for all given circuit depths $m$, each with a different layer $\mathsf{V}\in\mbb{L}(\mathsf{Clif}_1)$ sampled uniformly, preppended to it. We will denote $\mathsf{C}^*_m:=\mathsf{C}_m\mathsf{V}$.
    
    \item\label{protocol: quantum} \emph{Quantum Execution}: Estimate probabilities
    \begin{equation}
    \mathsf{P}_{\mathsf{W},\mathsf{C}_m^*}^{(\mathbf{s})} \! := \langle\mathbf{s}|\mathsf{W}\,\mathsf{C}_m\mathsf{V}(|0\rangle\!\langle0|)\mathsf{V}^\dg\mathsf{C}_m^\dagger\mathsf{W}^\dagger|\mathbf{s}\rangle,
    \end{equation}
    of observing the $n$-bit string $\mathbf{s}$, for all circuit depths $m$, all $N_{\mathsf{C}_m^*}$ circuit samples and a given number $N_\mathsf{W}$ of layer samples $\mathsf{W}\in\mbb{L}(\mbb{G}_1)$, sampled according to $\Omega_1$. We denote the corresponding estimator quantities, given a finite number of measurements $N_\mathsf{meas}$ by
    \begin{equation}
\hat{\mathsf{P}}_{\mathsf{W},\mathsf{C}_m^*}^{(\mathbf{s})}=\f{1}{N_\mathsf{meas}}\sum_{i=1}^{N_\mathsf{meas}}\boldsymbol{1}_\mathsf{s}(\mathbf{x}_i),
\label{eq: probability estimator}
    \end{equation}
    where $\mathbf{x}_i$ is a random variable describing the $i$\textsuperscript{th} projective measurement of the $n$-qubits in the computational basis, with $\mathbf{1}_\mathbf{s}(\mathbf{x})=1$ if $\mathbf{x}=\mathbf{s}$ or $0$ otherwise.
    
    \item\label{protocol: classical post-processing} \emph{Classical post-processing}: Estimate the average purity $\mathrm{tr}({\rho}_m^2)$ for all depth $m$ states $\rho_m:=\mathsf{C}_m\mathsf{V}(|0\rangle\!\langle0|)\mathsf{V}^\dg\mathsf{C}_m^\dagger$, via
    \begin{align}
        \hat{\mathfrak{P}}_m = 2^n\!\!\sum_{\substack{\mathbf{s},\mathbf{s}^\prime\\\mathsf{W},\mathsf{C}_m^*}}\!\f{(-2)^{-h(\mathbf{s},\mathbf{s}^\prime)} \hat{\mathsf{P}}^{(\mathbf{s})}_{\mathsf{W},\mathsf{C}_m^*}\hat{\mathsf{P}}^{(\mathbf{s}^\prime)}_{\mathsf{W},\mathsf{C}_m^*}}{N_{\mathsf{W}}N_{\mathsf{C}_m^*}},
        \label{eq: main average purity estimation in depth}
    \end{align}
    where $h(\mathbf{s},\mathbf{s})$ is the number of distinct elements between the $n$-bit strings $\mathbf{s}$ and $\mathbf{s}^\prime$. Estimate the probabilities $\hat{\mathsf{Q}}^{(\mathbf{s})}_{\mathsf{W},\mathsf{C}_m^{*}} := |\langle\mathbf{s}|\mathsf{W}^{(\mathsf{id})}|\psi^{(\mathsf{id})}_m\rangle|^2$ for all noiseless states, $\psi^{(\mathsf{id})}_m:=\mathsf{C}_m^{(\mathsf{id})}\mathsf{V}^{(\mathsf{id})}|0\rangle$ of depth $m$ ---with the label $(\mathsf{id})$ here denoting noiseless quantities---, and the corresponding average fidelities $\langle{\psi}_m|{\rho}_m|{\psi}_m\rangle$, via

    \begin{equation}
        \hat{\mathfrak{F}}_m = 2^n\!\!\!\sum_{\substack{\mathbf{s},\mathbf{s}^\prime\\\mathsf{W},\mathsf{C}_m^*}}\!\!\f{(-2)^{-h(\mathbf{s},\mathbf{s}^\prime)} \hat{\mathsf{P}}^{(\mathbf{s})}_{\mathsf{W},\mathsf{C}_m^*}\hat{\mathsf{Q}}^{(\mathbf{s}^\prime)}_{\mathsf{W},\mathsf{C}_m^*}}{N_{\mathsf{W}}N_{\mathsf{C}_m^*}}.
         \label{eq: main average fidelity estimation in depth}
    \end{equation}
    
    Both estimators can be rendered unbiased and improved by other considerations as in \S~\ref{sec: unbiased and MoMs}.
\end{enumerate}

Whenever the $\Omega$-distributed circuits are generated by layers containing non-Clifford gates in a way that becomes impractical to simulate (e.g., whereby there is a high density of such non-Cliffords or for a large number of qubits), the average layer fidelity can alternatively be estimated through \gls{mrb}~\cite{hines2022demonstrating}; the trade-off is having to construct and measure corresponding mirror circuits. The estimations of Protocol~\ref{sec: protocol} and \gls{mrb} by definition will agree upon employing the same ensemble of layers; in Appendix~\ref{appendix: numerical addendum} we numerically observe such agreement.

%\subsection{Main Results}\label{sec: main result}
\begin{result}[\emph{Exponential decay}]\label{result: exponential decay}Throughout Appendices~\ref{appendix: average purity of random circs},~\ref{appendix: operational estimation of the purity} and~\ref{appendix: operational estimation of fidelity} we show that, under the circumstances detailed below,
\begin{equation}
    \expect_\Omega\mathrm{tr}(\rho_m^2) \approx A\,\overline{\unitarity}(\mc{E})^m + \f{1}{2^n},
    \label{eq: main exponential decay}
\end{equation}
where $0\leq{A}\leq1$, and 
\begin{align}
\overline{\mathsf{u}}(\mc{E}) &= \f{2^n\overline{\gatefid}(\mc{E}^\dg\mc{E})-1}{2^n-1} = \overline{f}(\mc{E}^\dg\mc{E})
\label{eq: sq noise average}
\end{align}
is the average layer unitarity of the noise, where $\mc{E}:=\mc{L}_{i_\mathsf{noisy}}\mc{L}_i^\dg$ is approximately the noise channel corresponding to any layer $\mathsf{L}_i$ with associated map $\mc{L}_i(\cdot):=\mathsf{L}_i(\cdot)\mathsf{L}_i^\dg$, and $\overline{\gatefid}(\mc{E}^\dg\mc{E})$ is the average fidelity of $\mc{E}^\dg\mc{E}$ with respect to the identity.

Similarly, the decay in average fidelity can be seen to correspond to
\begin{align}
    \expect_\Omega\langle\psi_m|{\rho}_m|\psi_m\rangle \approx \alpha\overline{f}(\mc{E})^m + \f{1}{2^n},
    \label{eq: fidelity decay}
\end{align}
where $0\leq\alpha\leq1$, similar to a usual \gls{rb} decay for unital \gls{spam}.

In such case, the estimators in Eq.~\eqref{eq: main average purity estimation in depth} and Eq.~\eqref{eq: main average fidelity estimation in depth} of protocol~\ref{sec: protocol}, can be fit to the respective decays in depth $m$, whereby both average layer unitarity and fidelity can be estimated.
\end{result}

\subsubsection{Conditions for an exponential decay}\label{sec: conditions for exponential}
While a general functional form in terms of \emph{any} Markovian noise model for the average sequence purity and fidelity can be obtained, as in Appendices~\ref{appendix: average purity of random circs},~\ref{appendix: operational estimation of the purity} and~\ref{appendix: operational estimation of fidelity}, ensuring that it will follow a simple exponential decay as in Eq.~\eqref{eq: main exponential decay} relies on the noise satisfying certain approximate conditions and the pair $(\mbb{G},\Omega)$ having certain properties, similar to any \gls{rb}-based technique. Aside from standard assumptions such as Markovianity (i.e., that noise is not temporally correlated), time-independence and weak gate-dependence~\footnote{ Here meaning with time- and gate-dependent contributions being negligible on average; see e.g.,~\cite{characterizing_Magesan_2012}.}, Result~\ref{result: exponential decay} requires the following:

\begin{enumerate}[nolistsep, label=\subscript{A}{{\arabic*}}.]
    \item\label{assumption: tp unital} That noise is approximately trace-preserving and unital, i.e., without significant leakage or entropy-decreasing contributions.
    
    \item\label{assumption: 2-design} That the $\Omega$-distributed circuits approximate a unitary 2-design, to the effect that $\expect_{\mathsf{L}\sim\Omega}\mc{L}^\dg\mc{X}\mc{L}(\cdot)\approx{p}(\cdot)+(1-p)\mbb1/2^n$, where $p\leq1$ only dependent on $\mc{X}$; i.e., the average composition of noisy layers is approximately equivalent to a depolarizing channel encoding some information about $\mc{X}$. This is formalized and discussed in Appendix~\ref{appendix: unitary designs}.
\end{enumerate}

Assumption~\ref{assumption: tp unital} implies that the purity decay only depends on the unitarity of the average noise, and not on its non-unital or trace-decreasing contribution. Otherwise, if this condition is not satisfied, the average unitarity will not be described by Eq.~\eqref{eq: sq noise average}, but rather the most general Eq.~\eqref{eq: def uniform avg unitarity}, and the purity decay will be a convex combination of average unitarity and trace-decreasing contributions of the average noise, as expressed by Eq.~\eqref{eq: general purity decay non-unital non-TP} of Appendix~\ref{appendix: average purity of random circs} (in agreement with~\cite{Wallman_2015}). The implications for the fidelity estimation in this case are similar, since the polarization factor would not be described by Eq.~\eqref{eq: average polarization} but rather a more general $\overline{f}_t(\mc{E})=(2^n\overline{\gatefid}(\mc{E})-t/(2^n-1)$ for $t=\tr[\mc{E}(\mbb1/2^n)]$ quantifying how non-\gls{tp} the channel is~\cite{Wallman2018randomized}. In other words, the main implication is that the fitting procedure would generally be slightly more complicated for non-unital, non-\gls{tp} noise.

Assumption~\ref{assumption: 2-design} implies that we can identify $\overline{\mathsf{u}}(\mc{E})=p^2$ as given by Eq.~\eqref{eq: sq noise average}, i.e., the average unitarity is simply equal to the polarization of the average composition of noisy layers. This assumption is slightly stronger than that in~\cite {mrb_prl2022, hines2022demonstrating, hines2023fully, directRB2023}, and generally means that up to the second moment and a small positive $\epsilon$, the $\Omega$-distributed circuits we employ should reproduce the statistics of the uniform Haar measure on the unitary group on the $n$ qubits. While the choice of $(\mbb{G}_1,\Omega_1)$ being the uniform single-qubit Clifford group already generates separately a unitary 2-design on each qubit (projecting noise to a Pauli channel~\cite{simRB_2012}), the choice $(\mbb{G}_2,\Omega_2)$ should be such that the random $\Omega$-distributed circuits are \emph{highly scrambling}, as defined in~\cite{hines2022demonstrating, hines2023fully}, which can be achieved by it containing at least an entangling gate with high probability. While this does not ensure that the circuits will approximate a unitary 2-design in an increasing number of qubits, it suffices in practice for \emph{mid-scale systems}. This is discussed in detail in Appendix~\ref{appendix: unitary designs}.

\subsection{Mid-scale}
In \S~\ref{sec: scaling bottlenecks}, we will discuss the reasons why Protocol~\ref{sec: protocol} is practical and feasible for \emph{at least} 10-qubits and generally within tenths of qubits, which is what we refer to as \emph{mid-scale}. This is in the sense that all classical aspects, i.e., steps~\ref{protocol: preparation} and~\ref{protocol: classical post-processing} can be managed with a standard current-technology laptop and with the quantum execution requiring a mild number of circuits, \gls{rm}s, and measurement shot samples, as exemplified in \S~\ref{sec: demonstration}.

\subsection{Unbiased and Median of Means estimators}\label{sec: unbiased and MoMs}
There are at least two simple but effective ways in which reliable results can be ensured through the estimators in Eq.~\eqref{eq: main average purity estimation in depth} and Eq.~\eqref{eq: main average fidelity estimation in depth}: \emph{i}) using unbiased estimators~\footnote{ An estimator of some statistical parameter is said to be \emph{unbiased} or \emph{faithful} when its expectation matches the real expected value of the parameter, or otherwise it is said to be biased.} and \emph{ii}) using \gls{moms} estimators.

In~\cite{Zoller_2018} it is pointed out that, even though the probability estimator in Eq.~\eqref{eq: probability estimator} is faithful, i.e. $\expect[\hat{\mathsf{P}}]=\mathsf{P}$ (where we have dropped all indices), it is biased for any other positive integer power. A unique unbiased estimator can nevertheless be built for any power, and in particular, $\hat{\mathsf{P}}_2:=\hat{\mathsf{P}}\left(\hat{\mathsf{P}}N_\mathsf{meas}-1\right)/\left(N_\mathsf{meas}-1\right)$ is an unbiased estimator of $\mathsf{P}^2$. Thus all terms in Eq.~\eqref{eq: main average purity estimation in depth} and Eq.~\eqref{eq: main average fidelity estimation in depth} for the $n$-bit string $\mathbf{s}=\mathbf{s}^\prime$ should be computed through $\hat{\mathsf{P}}_2$ to render average sequence fidelity and purity estimations unbiased.

On the other hand, a simple but highly effective way to reduce the uncertainty associated to a mean estimator is to use \gls{moms} estimators: given $K$ samples of estimators $\hat{\mathfrak{F}}_m^{(1)},\cdots,\hat{\mathfrak{F}}_m^{(K)}$ for the average fidelity of circuits of depth $m$ in Eq.~\eqref{eq: main average fidelity estimation in depth}, the \gls{moms} estimator of such samples is $\hat{\mathfrak{F}}_m^{\mathsf{MoMs}(K)}:=\mathsf{median}(\hat{\mathfrak{F}}_m^{(1)},\cdots,\hat{\mathfrak{F}}_m^{(K)})$, or similarly for the case of the average purity in Eq.~\eqref{eq: main average purity estimation in depth}. In total, this requires $N_\mathsf{W}N_{\mathsf{C}_m^*}K$ circuit samples to be measured $N_\mathsf{meas}$ times, which however, is generally more robust compared with simply constructing an empirical estimator of the mean with the same amount of samples~\cite{shadows_2020}. While it has been noted that when measurements are randomized with the single-qubit Clifford group both types of estimators converge similarly to the true mean~\cite{PhysRevLett.127.110504}, \gls{moms} has the concentration property of the probability of observing outliers from the true mean decreasing exponentially in $K$~\cite{lerasle2019lecture}.

\section{Experiment on \texorpdfstring{\spark}{IQM Spark (TM)}}\label{sec: demonstration}
We now demonstrate the execution of Protocol~\ref{sec: protocol} through both an experiment on \spark~\cite{spark}, a commercial 5-qubit superconducting quantum system by IQM targeting education in quantum computing; technical details of the hardware can be seen in~\cite{spark_paper}. The five qubits in the \spark~are connected in a star-shape, with a central qubit connected to the four remaining qubits. The experiments were performed at specific dates, pointed out where the respective results are displayed, and only reflect the performance of the hardware at such point in time.

\begin{figure}[t!]
    \centering \includegraphics[width=0.5\textwidth]{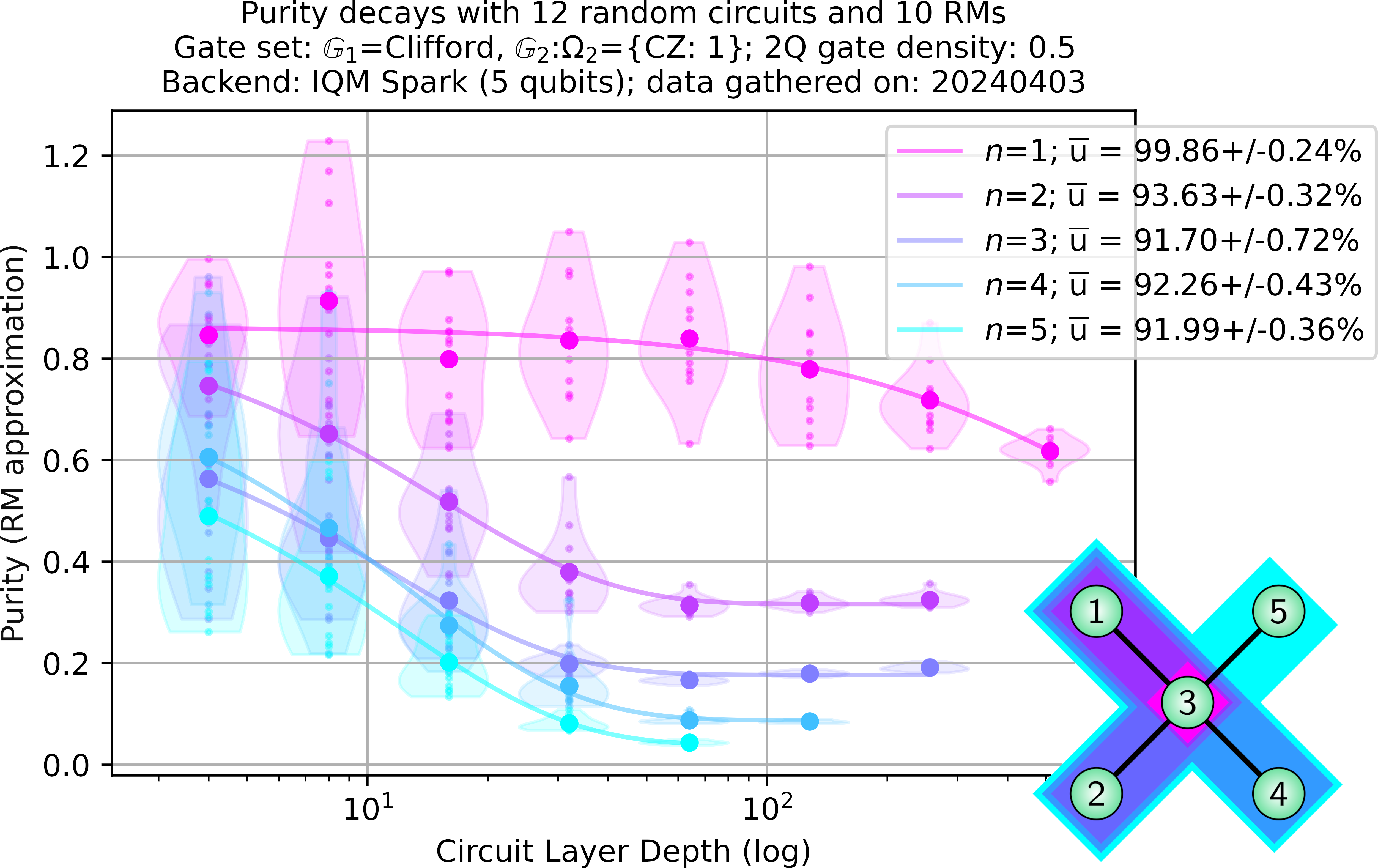}
    \caption{\textbf{Purity decays on \spark}. Computed according to Protocol~\ref{sec: protocol} for $\Omega$-distributed circuits with gate set $\mbb{G}=\mathsf{Clif}_1\cup\{\mathsf{CZ}\}$, with $\mathsf{Clif}_1$ being the uniformly-distributed single-qubit Clifford group. Individual plots correspond to a given number of qubits taken from a connectivity graph shown in the inset, whereby smaller points correspond to \gls{moms} estimators of purity in Eq.~\eqref{eq: main average purity estimation in depth} at a given depth $m$, bigger points correspond to the average purity, violins show the distribution of the \gls{moms} estimators, and lines correspond to a least-squares fit of the averages to the model in Eq.~\eqref{eq: main exponential decay}, with which the corresponding average unitarities $\overline{\unitarity}$ are estimated. Sample parameters: $N_{\mathsf{C}_m^*}=12$, $N_\mathsf{W}=10$, $N_{\mathsf{meas}}=2^{11}$ and $K=1,2$ (for $n=1,2,3$ and $n=4,5$ respectively)  \gls{moms}.}
    \label{fig: purity spark}
\end{figure}

We generated $\Omega$-distributed circuits using the gate set $\mbb{G}=\mathsf{Clif}_1\cup\{\mathsf{CZ}\}$, with $\mathsf{Clif}_1$ being the uniformly-distributed single-qubit Clifford group (i.e., $\Omega_1$ assigning 1/24 probability for each Clifford and $\Omega_2$ probability 1 of sampling $\mathsf{CZ}$). The sampling of layers was done employing the edge-grab sampler, defined in~\cite{Proctor_2021}, with a two-qubit gate density (the expected proportion of qubits occupied by two-qubit gates) of 1/2. The edge-grab sampler considers layers with two-qubit gates only on connected qubits, and where a single logical layer consists of a parallel mixture of one- and two-qubit gates (sampled according to $\Omega$), i.e., there is no more than one gate acting on a given qubit in any layer.
 
The native set of gates is $\{r_{\theta,\varphi},\mathsf{CZ}\}$, where $r_{\theta,\varphi}$ is a single-qubit rotation of angle $\theta$ around the $\cos(\varphi)X + \sin(\varphi)Y$ axis, $\mathsf{CZ}$ is a controlled-$Z$ two-qubit gate. The $\mathsf{measurement}$ operator is a computational basis projective measurement operation, and a $\mathsf{barrier}$ object is used to prevent the layers defined through $\mathbb{G}$ from being compiled together. The definition of \emph{depth} that we adopt is that of the number of layers defined through $\mathbb{G}$, i.e., before transpilation to the native gate set.

We extracted decay rates of the average state fidelity and purity in increasing layer depths according to Protocol~\ref{sec: protocol}, and thus layer fidelity and unitarity, for $\Omega$-distributed circuits in arbitrary combinations of $n=1,2,\ldots,5$ qubits. In both cases, the sample parameters we employ are $N_{\mathsf{C}_m^*}=12$ random $\Omega$-distributed circuits, times $N_\mathsf{W}=10$ randomized measurement layers, and times $N_{\mathsf{meas}}=2^{11}$ shots per measurement; we furthermore employ unbiased estimators for square probabilities and either $K=1$ for $n=1,2,3$ or $K=2$ for $n=4,5$ \glsreset{moms}\gls{moms}\glsreset{moms} estimators.

\begin{figure}[t!]
    \centering
    \includegraphics[width=0.5\textwidth]{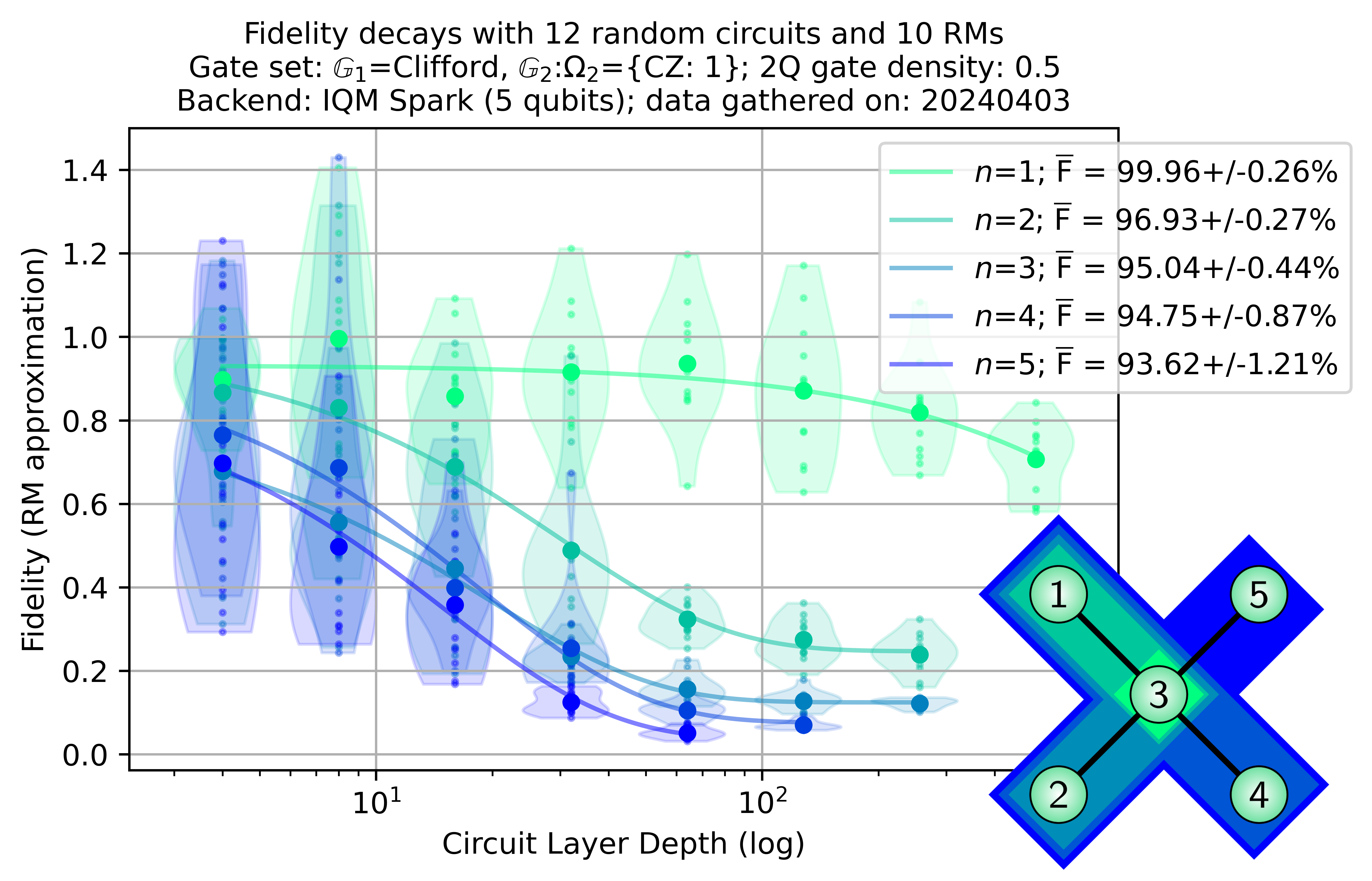}
    \caption{\textbf{Fidelity decays on \spark}. Fidelity decays according to Protocol~\ref{sec: protocol} corresponding to the same $\Omega$-distributed circuits and parameters giving probabilities with purities displayed in Fig.~\ref{fig: purity}; here similarly individual plots correspond to a given number of qubits $\mathsf{n}$, small points correspond to \gls{moms} estimators of~\eqref{eq: main average fidelity estimation in depth} at the given depth, larger points correspond to the respective average purity, and lines correspond to least-squares fit of the averages to the model in Eq.~\eqref{eq: fidelity decay}, whereby the corresponding average layer fidelity $\overline{\gatefid}$ in the respective number of qubits $\mathsf{n}$ is extracted according to Eq.~\eqref{eq: average polarization}.}
    \label{fig: fidelity spark}
\end{figure}

In Fig.~\ref{fig: purity spark} we show the purity decays in increasing circuit depth for each number of qubits, determined via the estimators in Eq.~\eqref{eq: main average purity estimation in depth}, where, by fitting the decay model in Eq.~\eqref{eq: main exponential decay} to the averages, we estimate the corresponding average unitarity. It is worth noticing that we used no median of means for up to three qubits, and thus the individual smaller points in the plot do not represent purities as these have not been averaged over all circuit samples; this is a reason why the violin distributions can stretch beyond 1. Nevertheless, it is expected for the spread in individual outputs to be larger for larger values of purity. Importantly, while averages (larger points) do not all fall exactly on the respective exponential (particularly for $n=1$), deviations do not appear alarmingly high. Finally, \gls{spam} contributions can be seen to lead to a shift in the offset of the curve for $n=3$.

It is important to notice that in Fig.~\ref{fig: purity spark}, the unitarity does not appear to be monotonically decreasing in qubit count. In this case, however, this apparent increase could also be explained away by the uncertainty of the data and could be at most a plateau in unitarity; a way to decrease the uncertainties could be by increasing the number of \gls{moms}. This is nevertheless an interesting feature and one that could be investigated further also in the context of determining crosstalk or other types of correlations at the unitarity level.

\begin{figure}[t!]
    \includegraphics[width=0.5\textwidth]{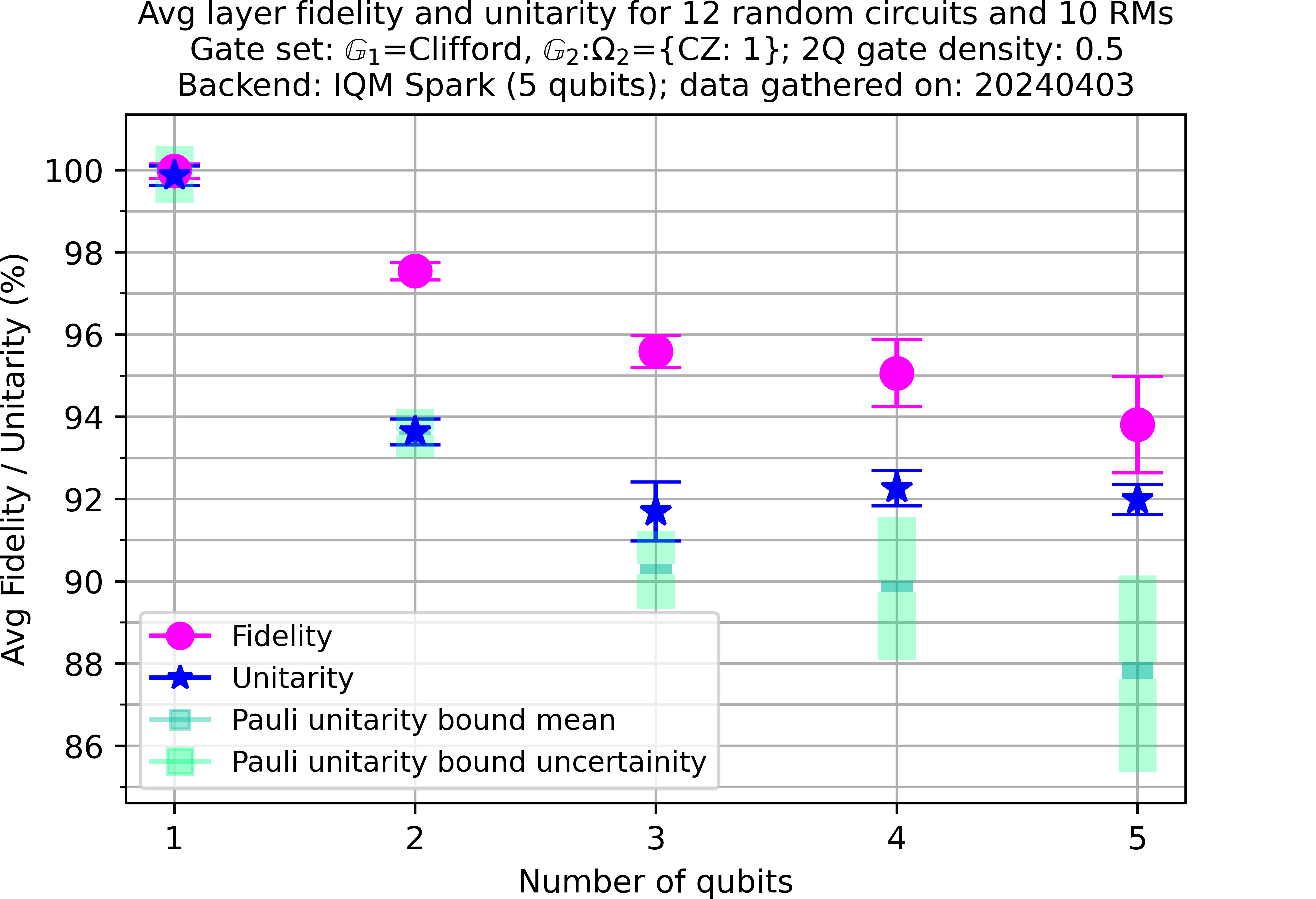}
    \caption{\textbf{Average layer fidelities and average unitarities on \spark}. All estimated average layer fidelities and unitarities, extracted experimentally through Protocol~\ref{sec: protocol}, in increasing number of qubits are shown, together with ranges of values for potential Pauli unitarities, computed with the estimated fidelities according to Ineq.~\eqref{eq: main Pauli unitarity bounds}.}
    \label{fig: unitarity and fidelity spark}
\end{figure}

Following this, in Fig.~\ref{fig: fidelity spark} we show the respective fidelity decays, computed together with the measurement counts from the same experiment and the estimation of the corresponding noiseless probabilities, according to estimator in Eq.~\eqref{eq: main average fidelity estimation in depth}. Similarly here, the distributions tend to show a larger spread for smaller $n$, where only the averages represent the corresponding fidelity estimation. While the decays display different \gls{spam} contributions, the averages do decrease in system size; the uncertainties could similarly be reduced e.g., by increasing the amount of \gls{moms}. In Appendix~\ref{appendix: mrb comparison spark} we compare with \gls{mrb} results gathered weeks before, showing qualitative agreement despite a clear difference in the respective distributions. Since error bars remain relatively large for fidelities, and hence for the estimated Pauli unitarity intervals, an alternative could be to construct mirror circuits (without the \gls{rm} Clifford gates appended) within the same experiment and then instead estimate fidelities via \gls{mrb} for smaller uncertainties.

This finally enables us to get an overall picture of both average layer fidelities and unitarities, together with an estimation of whether noise falls within the Pauli unitarity bound in Eq.~\eqref{eq: main Pauli unitarity bounds} or it contains a larger coherent contribution. In Fig.~\ref{fig: unitarity and fidelity spark} we plot together the average outputs retrieved in Fig.~\ref{fig: purity spark} and Fig.~\ref{fig: fidelity spark}, together with estimations of Pauli unitarity bounds given the average fidelity estimates, computed via Eq.~\eqref{eq: main Pauli unitarity bounds}. Fig.~\ref{fig: unitarity and fidelity spark} gives an overall picture of a coherent noise budget, which could in turn inform the overhead of Pauli twirling techniques. It is relevant to notice that both it is possible for the average unitarity to increase even without there being necessarily correlations between qubits, and that generally the coherence of noise could remain relatively large with respect to fidelity for larger systems.

\section{Scaling bottlenecks}\label{sec: scaling bottlenecks}
\subsection{Sample Complexity}\label{sec: sample complexity}
The main roadblock for estimating the average coherence of noise in large systems (or even just the purity of quantum states) with \gls{rm}s, is the required number of measurements to have an estimation within a given error, i.e., its sample complexity, which is then by definition manifested by the variance associated to such \gls{rm}s. In Protocol~\ref{sec: protocol}, statistical uncertainty stems from the sampling of the $\Omega$-distributed circuits, that of the \gls{rm} elements, and the number of measurements per \gls{rm} element; these quantities correspond to $N_{\mathsf{C}_m^*}$ and $N_\mathsf{W}$ in the estimators of Eq.~\eqref{eq: main average fidelity estimation in depth} and Eq.~\eqref{eq: main average purity estimation in depth}, and to $N_\mathsf{meas}$ of Eq.~\eqref{eq: probability estimator}, respectively.

The main limitation to scalability is imposed by $N_\mathsf{meas}$: using local \gls{rm}s via Eq.~\eqref{eq: purity random cross-correlations} to estimate purity to precision $1/\sqrt{N_\mathsf{W}}$, it was observed numerically in~\cite{purity_pra_2019} that it scales approximately as $N_\mathsf{meas}\sim{2}^{0.75n}$ in number of qubits $n$. Generally, within the classical shadows framework, the required number of measurements can be seen to scale exponentially in a similar fraction of $n$ and linearly in the actual purity of the state~\cite{shadows_2020, PhysRevLett.125.200501, vermersch2023manybody}.

Mixedness reduces the sample complexity, as also observed numerically in~\cite{purity_pra_2019}, and with $\Omega$-distributed circuits, quantum states will naturally get increasingly mixed in increasing depth. Coincidentally, thus, a high percentage of coherent error contributes to a higher variance not only in average fidelity estimation but also in average unitarity estimation itself.

When it comes to $\Omega$-distributed circuit samples when estimating average fidelity, the variance associated with it can be directly linked to the average amount of coherent noise~\cite{Wallman_2015, hines2022demonstrating}. How to control this uncertainty, as well as guarantees when choosing the number of sample circuits and number of measurements, has been widely studied for Clifford \gls{rb}~\cite{characterizing_Magesan_2012, PhysRevA.100.032304, Wallman_2014}; furthermore, \cite{unitarity_helsen_2019} obtained a bound on the variance for the purity in circuits within (an improved version of) Clifford \gls{urb}, assuming unital noise. While we are not aware of analogous bounds on the variance for scalable \gls{rb} (namely employing $\Omega$-distributed circuits), it has been observed numerically that similar behavior follows (at least when layers are made up only of either generators-of or Clifford gates themselves)~\cite{hines2022demonstrating, hines2023fully}, and that \emph{Pauli twirling}~\cite{PhysRevA.103.042604, PhysRevA.94.052325} generally suppresses the variance of expectation values under any sequence of operations, regardless of whether noise is correlated (either spatially or temporally) or not~\cite{figueroaromero2023operational}.

A prospect could therefore be to benchmark average Pauli-dressed gates, aiming to obtain a unitarity within Ineq.~\eqref{eq: main Pauli unitarity bounds}, with the trade-off being to sample an extra number of Pauli layers per circuit. Here we do not attempt to derive a \emph{rigorous} bound on the variance for the average layer unitarity, but as we have demonstrated experimentally in \S~\ref{sec: demonstration}, while for a given total number of samples the uncertainty is larger for the unitarity than for fidelity, estimating it nevertheless remains tractable for \emph{mid-scale} systems.

\subsection{Practical bottlenecks}
There are two other \emph{practical} considerations to take into account to estimate at which scale Protocol~\ref{protocol: classical post-processing} remains feasible: the \gls{spam} factors in the exponential decay in Eqs.~\eqref{eq: main exponential decay},\eqref{eq: fidelity decay}, and the number of terms entering in the sum of Eqs.~\eqref{eq: main average purity estimation in depth},\eqref{eq: main average fidelity estimation in depth}.

While our technique gives \gls{spam}-independent estimates of the average unitarity, it can nevertheless make the exponential decays drop way too quickly, so that fitting becomes unfeasible. This is a shared problem of all \gls{rb}-based techniques, and other than simply improving readouts and/or state preparation, there could be ways of removing the $2^{-n}$ offset, for example, but the multiplicative factor is still determined by the unitarity of the \gls{spam}~\footnote{ For example, if we assume that \gls{spam} noise (i.e. that which we would attribute to the initial and final single-qubit-only layers) is global depolarizing (as done for \gls{mrb}~\cite{hines2022demonstrating}) with polarization $p$, then the multiplicative factor in Eq.~\eqref{eq: main exponential decay} is $A\lesssim{p}^4$.}. More than a \gls{rb} problem this might be an issue in general when trying to estimate global, high-weight properties on large systems, essentially because signal readouts eventually become too small.

On the other hand, while Eq.~\eqref{eq: purity random cross-correlations} and Eq.~\eqref{eq: probability purity orig} are remarkable mathematically, in practice they involve a sum over all possible pairs of $n$-bit strings. Naively, this means summing up $4^n$ probability terms; for the case of the purity, this can be reduced at least to $2^{n-1}(2^n-1)$, given that the Hamming distance is symmetric. This computation can be done easily on a classical machine in parallel, but it can also become rather unpractical (as it furthermore would need to be done for every single purity estimate). While this is a practical problem, it is fundamentally tied to the fact that the purity of a state involves all the elements of the density matrix.

\section{Conclusions}\label{sec: conclusions}
\glsresetall
We have derived an upper bound on the average unitarity of Pauli-twirled noise solely in terms of the average fidelity of the respective bare noise, and we have established a protocol enabling the estimation of the average unitarity for a broad class of quantum circuits in digital systems, independent of technology platform or architecture, with tenths of qubits. Our results have been inspired both by novel scalable \gls{rb} techniques, as well as \gls{rm} methods. We have shown that reliable estimation of both average operation coherence and fidelity in such systems can be done under mild conditions and with a reasonable sampling overhead. Finally, we demonstrated our results in experiment on up to 5 qubits and numerically in simulation with up to 10 qubits.

A fully scalable estimation of the average coherence of noise will almost surely rely on modularity, e.g., by using smaller-scale unitarity information extracted from smaller qubit subsets and then combined with a Simultaneous \gls{rb}-like protocol.

A promising way of moving forward in this direction is that of~\cite{vermersch2023manybody}, where the authors establish a method to estimate entropy and entanglement of a quantum state in polynomially-many measurements, relying on conditions essentially satisfied when spatial correlation lengths in the system remain finite and for systems in a one-dimensional topology. Their main result suggests that generally one could partition \emph{any} large $n$-qubit system into smaller subsystems and reconstruct the global $n$-qubit purity from the \emph{local} purity of pairs of contiguous subsystems, like \emph{stitching} together the bigger purity with a number of measurements that overall grow polynomially in $n$. As a direct consequence, if such results hold more generally, our protocol could be used to estimate global average noise coherence with a similar sample complexity.

Nevertheless, several things remain unclear, e.g., what classes of (Markovian) noise such a method would hold for, how it could be phrased beyond one-dimensional qubit arrays, or how one would certify the global purity estimations, among others. As we have argued, however, the average coherence of noise is a crucial figure of merit for benchmarking the error mechanisms limiting the performance of quantum computers, perhaps only standing after average layer fidelity in terms of importance as a benchmark, so the aforementioned hurdles must soon be overcome.

\begin{acknowledgments}
The authors acknowledge support from the German Federal Ministry of Education and Research (BMBF) under Q-Exa (grant No. 13N16062) and QSolid (grant No. 13N16161). The authors also acknowledge the entire IQM Technology team for their support in the development of this work.
\end{acknowledgments}
\vfill

\bibliographystyle{apsrev4-1_custom}

\appendix
\addtocontents{toc}{\protect\setcounter{tocdepth}{0}}
\onecolumn
\glsresetall

\section{Preliminaries}
Here we detail and expand on some of the notation and technical details that go into showing the results claimed in the main text. Much of the structure for the circuits entering the protocol is inspired by~\cite{mrb_prl2022, hines2022demonstrating}, so for this reason we stick to most of the original terminology and notation.

\subsection{\texorpdfstring{$\Omega$}{Omega}-distributed circuits}
We will consider a quantum system made up of $n\geq2$ qubits and a pair of single-qubit and two-qubit gate sets, $\mbb{G}_1$ and $\mbb{G}_2$, respectively, with corresponding probability distributions $\Omega_1$ and $\Omega_2$. We will then generate quantum circuits by sampling from $\mathsf{g}_1\sim\Omega_{1}$ and $\mathsf{g}_2\sim\Omega_{2}$, to generate parallel instructions on the $n$ qubits as specified by a set $\mbb{L}(\mbb{G})=\{L_i=L_i^{(1)}L_i^{(2)}\}$, where each $L_i^{(1)}\in\mbb{L}(\mbb{G}_1)$ and $L_i^{(2)}\in\mbb{L}(\mbb{G}_2)$ are instructions made up of only the single and two-qubit gates $\mathsf{g}_1$ and $\mathsf{g}_2$, respectively. To stress when some quantities are random variables, we use sans font when relevant, as in $\mathsf{L}_i$ being a random layer sampled from $\Omega$, which is such that $\Omega(\mathsf{L}_i)=\Omega_1(\mathsf{L}_i^{(1)})\Omega_2(\mathsf{L}_i^{(2)})$.

Circuits generated this way, here for the particular case $\mbb{G}=(\mbb{G}_1,\mbb{G}_2)$ and $\Omega=(\Omega_1,\Omega_2)$, are said to be $\Omega$-distributed circuits, e.g., the circuit as stated in the main text as
\begin{equation}
    \mathsf{C}_m=\mathsf{L}_m\mathsf{L}_{m-1}\cdots\mathsf{L}_1,
\end{equation}
where each $\mathsf{L}_i\in\mbb{L}(\Omega)$ sampled according to $\Omega$, and the right-hand-side is read from right to left (i.e., $\mathsf{C}_m$ is an operator acting onto quantum states on the left), is a $\Omega$-distributed circuit of depth $m$.

We point out that, in general, $\Omega$-distributed circuits can be constructed with an arbitrary number of gate sets and their corresponding distributions. Moreover, while so far $\mbb{G}_1,\mbb{G}_2$ and $\Omega_1,\Omega_2$ are completely up to the user to specify, in practice the main requirements we will have on this choice are that, \emph{i}) the resulting $\Omega$-distributed circuits constitute an $\epsilon$-approximate unitary 2-design, and \emph{ii}) that in particular $(\mbb{G}_1,\Omega_1)$ is an exact unitary 2-design, e.g., the single-qubit Clifford group. As such conditions \emph{i}) and \emph{ii}) are considered \emph{a posteriori} for a particular case of interest, we will have a general derivation and then these will be stated formally in point~\ref{appendix: purity to exponential} of \S~\ref{appendix: average purity m steps} and in \S~\ref{appendix: purity with rms}, respectively.

\subsection{Ideal vs real implementations}
While we do not stress this difference in the main text, here we will denote by $U_L$ an ideal (noiseless) unitary operator associated to a representation of the layer $L$, and by $\mc{U}_L(\rho):=U_L\rho U_L^\dg$ the action of its corresponding map or superoperator $\mc{U}_L$.

We write a composition of two maps $\mc{A}$ and $\mc{B}$ simply as $\mc{B}\mc{A}$ to mean ``apply map $\mc{B}$ after applying map $\mc{A}$'', equivalent to the usual notations $\mc{B}\circ\mc{A}$, or $\mc{B}(\mc{A}(x))$.

Noisy implementations of layers will generally be denoted by a map $\phi$, e.g., the noisy application of a layer $L$ is denoted by the \gls{cp} map (which we also refer to as a quantum channel, or simply, a channel) $\phi(L)$, which can be defined (see e.g., \S{IV} of~\cite{PRXQuantum.3.020335}) such that $\phi(L):=\mc{E}_L\,\mc{U}_L$, with $\mc{E}_L$ being generally a \gls{cp} map associated to $L$.

\subsection{The output states of \texorpdfstring{$\Omega$}{Omega}-distributed circuits}
We will consider a set of $\Omega$-distributed circuits of depth $m$, acting on a fiducial initial state $|0\rangle$, which in turn we randomize with a layer $\mathsf{V}\in\mbb{L}(\mbb{G}_1)$. That is, we will have circuits of the form $\mathsf{C}_m\mathsf{V}|0\rangle$, which will always give a pure state output in the noiseless case. In the noisy case, however, we will have states of the form
\begin{equation}
    \tilde{\varrho}_m := \expect_{\substack{\mathsf{L}_i\sim\Omega\\\mathsf{V}\sim \Omega_1}}\phi(\mathsf{L}_m)\phi({\mathsf{L}_{m-1}})\cdots\phi(\mathsf{L}_1)\phi(\mathsf{V})(|0\rangle\!\langle{0}|),
    \label{eq: avg state layer m}
\end{equation}
averaged independently over each of the $\Omega$-distributed layers $\mathsf{L}_i$, and uniformly over the initial single-qubit gate layer $\mathsf{V}\in\mbb{L}(\mbb{G}_1)$ with $\Omega_1$. This noisy output can of course now be mixed, $2^{-N}\leq\tr(\tilde{\varrho}_m^2)\leq1$.

Our protocol focuses on efficiently estimating the purities of the states $\tilde{\varrho}_m$, to in turn estimate the average unitarity of noise in the circuits $\mathsf{C}_m$. In the following, we first study the behavior of the average purity of $\tilde{\varrho}_m$ in increasing circuit depth $m$, and then we see that we can efficiently estimate this quantity employing \gls{rm}s, allowing us to readily extract the average unitarity of noise in certain circumstances which we analyze.

\section{Purity of average outputs of \texorpdfstring{$\Omega$}{Omega}-distributed circuits}\label{appendix: average purity of random circs}
\subsection{Purity of the initial randomization}
Let us consider first the case of no gate layers but just the noisy initial single-qubit Clifford layer, $m=0$, then we have
\begin{align}
    \tr(\tilde{\varrho}_0^2) &= \expect_{\mathsf{V}\sim\Omega_1} \tr[\phi(\mathsf{V})(|0\rangle\!\langle0|)\phi(\mathsf{V})(|0\rangle\!\langle0|)^\dg] \nonumber\\
    &:= \expect_{\psi\sim\Omega_1}\tr[\mc{E}_\mathsf{spam}(|\psi\rangle\!\langle\psi|)\mc{E}_\mathsf{spam}(|\psi\rangle\!\langle\psi|)^\dg],
\end{align}
where we defined the noisy implementation of $\mathsf{V}$ as $\phi(\mathsf{V}):=\mc{E}_\mathsf{spam}\,\mc{U}_{\mathsf{V}}$ for some noise channel $\mc{E}_\mathsf{spam}$ explicitly standing for \gls{spam} noise, and where we defined the initial randomized pure state as $\mc{U}_{\mathsf{V}}(|0\rangle\!\langle0|):=|\psi\rangle\!\langle\psi|$.

Noticing that the noisy outputs $\mc{E}_\mathsf{spam}(|\psi\rangle\!\langle\psi|)$ are quantum states and thus Hermitian, and that we can define the self-adjoint map $\mc{X}^\dg$ of a quantum channel $\mc{X}$ by the property $\tr[A\mc{X}(B)]:=\tr[\mc{X}^\dg(A)B]$, we can write
\begin{align}
    \tr(\tilde{\varrho}_0^2) &= \expect_{\psi\sim\Omega_1}\langle\psi|\mc{E}_\mathsf{spam}^\dg\mc{E}_\mathsf{spam}(|\psi\rangle\!\langle\psi|)|\psi\rangle.
    \label{eq: purity m=1 def}
\end{align}
Equivalently, the self-adjoint map of a quantum channel can be defined in terms of its Kraus operators as $\mc{X}^\dg(\cdot):=\sum{K}_\mu^\dg(\cdot)K_\mu$, where $\{K_\mu\}$ the Kraus operators of $\mc{X}$, so Eq.~\eqref{eq: purity m=1 def} in a sense is already measuring how different $\mc{E}_\mathsf{spam}$ is from a unitary.

In general, the purity of a quantum state, $\rho:=\mc{X}(|\varphi\rangle\!\langle\varphi|)$ can be written as $\tr(\rho^2):=\langle\varphi|\mc{X}^\dg\mc{X}(|\varphi\rangle\!\langle\varphi|)|\varphi\rangle$, which corresponds to the gate fidelity of $\mc{X}^\dg\mc{X}$ with respect to the identity on the state $|\varphi\rangle$. Thus, we can write Eq.~\eqref{eq: purity m=1 def} as a gate-fidelity of $\mc{E}_\mathsf{spam}^\dg\mc{E}_\mathsf{spam}$ with respect to the identity, averaged over all possible initial states $|\psi\rangle$,
\begin{align}
    \tr(\tilde{\varrho}_0^2)
    &= \expect_{\psi\sim\Omega_1} \gatefid_\psi(\mc{E}_\mathsf{spam}^\dg\mc{E}_\mathsf{spam}),
    \label{appendix eq: avg purity and fidelity}
\end{align}
where here 
\begin{equation}
    \gatefid_\psi(\mc{X}):=\langle\psi|\mc{X}(|\psi\rangle\!\langle\psi|)|\psi\rangle,
\end{equation}
is the gate-fidelity of the map $\mc{X}$ with respect to the identity map on the state $|\psi\rangle$.

\subsection{Relation to average unitarity and average trace-decrease}
Ultimately, however, the average loss of purity in the average state $\tilde{\varrho}_0$ must be related to some amount of \emph{non-unitarity} of the noise, i.e., it is noise that decreases purity. This can be quantified by the average unitarity, $\expect_{|\psi\rangle}\mathsf{u}_\psi$, defined for a \gls{cp} map $\mc{E}$ as
\begin{equation}
    \expect_{\psi\sim\Omega_1}\mathsf{u}_\psi(\mc{E}):=\left(\f{2^n}{2^n-1}\right)\expect_{\psi\sim\Omega_1}\gatefid_\psi(\mc{E}^{\prime\,\dg}\mc{E}^\prime),\quad\text{where}\quad\mc{E}^\prime(\cdot):=\mc{E}(\cdot-\mbb1/2^n),
\end{equation}
which is so defined to account for the case of $\mc{E}$ being trace-decreasing. When $\mc{E}$ is \gls{tp} we have the relation $\mc{E}^{\dg\,\prime}\mc{E}^\prime(\cdot)=\mc{E}^{\dg}\mc{E}(\cdot-\mbb1/2^n)$, otherwise, however, for any quantum state $\rho$,
\begin{equation}
    \mc{E}^{\dg\,\prime}\mc{E}^\prime(\rho) = \mc{E}^{\dg}\mc{E}\left(\rho-\f{\mbb1}{2^n}\right)-\tr\left[\mc{E}\left(\rho-\f{\mbb1}{2^n}\right)\right]\mc{E}^\dg\left(\f{\mbb1}{2^n}\right),
\end{equation}
thus
\begin{align}
    \left(\f{2^n-1}{2^n}\right)\expect_{\psi\sim\Omega_1}\unitarity_\psi(\mc{E}) &= \expect_{\psi\sim\Omega_1} \gatefid_\psi(\mc{E}^\dg\mc{E}) \nonumber\\
    &\qquad- \expect_{\psi\sim\Omega_1}\langle\psi|\mc{E}^\dg\mc{E}(\mbb1/2^n)|\psi\rangle - \expect_{\psi\sim\Omega_1}\tr\left[\mc{E}^\prime\left(|\psi\rangle\!\langle\psi|\right)\right]\langle\psi|\mc{E}^\dg(\mbb1/2^n)|\psi\rangle \nonumber\\
    &= \expect_{\psi\sim\Omega_1} \gatefid_\psi(\mc{E}^\dg\mc{E}) - 2^{-n}\expect_{\psi\sim\Omega_1}\left\{\mathsf{S}_\psi(\mc{E}^\dg\mc{E}) + \mathsf{S}_\psi(\mc{E})\left[\mathsf{S}_\psi(\mc{E}) - \tr[\mc{E}(\mbb1/2^n)]\right]\right\},
    \label{eq: unitarity and fid plus tp}
\end{align}
where
\begin{equation}
    \mathsf{S}_\psi(\mc{E}):=\tr[\mc{E}(|\psi\rangle\!\langle\psi|)],
\end{equation}
is a trace-preservation measure of the map $\mc{E}$ for a given sample $|\psi\rangle$.

We can then write the average purity of $\tilde{\varrho}_0$ as
\begin{align}
    \tr(\tilde{\varrho}_0^2) &= \left(\f{2^n-1}{2^n}\right)\expect_{\psi\sim\Omega_1}\unitarity_\psi(\mc{E}_\mathsf{spam}) \nonumber\\
    &\quad+ 2^{-n}\expect_{\psi\sim\Omega_1}\left\{\mathsf{S}_\psi(\mc{E}_\mathsf{spam}^\dg\mc{E}_\mathsf{spam}) + \mathsf{S}_\psi(\mc{E}_\mathsf{spam})\left[\mathsf{S}_\psi(\mc{E}_\mathsf{spam}) - \tr[\mc{E}_\mathsf{spam}(\mbb1/2^n)]\right]\right\},
    \label{eq: avg purity m=1 general spam}
\end{align}
which we recall is defined as averaged over the states $|\psi\rangle\!\langle\psi|=\mc{U}_\mathsf{V}(|0\rangle\!\langle0|)$, which are distributed uniformly and independently on the $n$-qubits according to a distribution $\Omega_1$ on single-qubit gates $\mbb{G}_1$.

When the noise channel $\mc{E}$ is \gls{tp}, the rightmost summand in Eq.~\eqref{eq: unitarity and fid plus tp} becomes a term equal to $-2^{-n}$, and in that case the average unitarity is just a re-scaling of the average gate-fidelity of the form
\begin{equation}
    \expect_{\psi\sim\Omega_1}\unitarity_\psi(\mc{E})\quad \stackrel{\scriptscriptstyle{(\mc{E}\,\text{is TP})}}{=}\quad \f{2^n \expect_{\psi\sim\Omega_1}\gatefid_\psi(\mc{E}^\dg\mc{E})-1}{2^n-1}.
\end{equation}
and thus also the average purity in Eq.~\eqref{appendix eq: avg purity and fidelity} in this case would become
\begin{align}
    \tr(\tilde{\varrho}_0^2) &\stackrel{\scriptscriptstyle{(\mc{E}_\mathsf{spam}\,\text{is TP})}}{=} \left(\f{2^n-1}{2^n}\right)\expect_{\psi\sim\Omega_1}\unitarity_\psi(\mc{E}_\mathsf{spam}) + \f{1}{2^n}.
\end{align}

\subsection{The case of global depolarizing noise}\label{appendix: the case of global depolarizing noise}
A particular case of a quantum channel for modeling noise that is both easy mathematically as in interpretation is that of global depolarizing channel,
\begin{align}
    \mc{E}_p(\cdot):=p\,(\cdot)+(1-p)\tr(\cdot)\,\f{\mbb1}{2^n},
\end{align}
where $0\leq{p}\leq1$ is the probability for the input state to remain the same, and which can be directly defined in terms of the channel by the so-called polarization $\gamma$,
\begin{equation}
    \gamma(\mc{E}) := \f{4^n\entfid(\mc{E})-1}{4^n-1},
\end{equation}
where here 
\begin{equation}
    \entfid(\mc{E}):=\langle\Psi|(\mc{I}\otimes\mc{E})[|\Psi\rangle\!\langle\Psi|]|\Psi\rangle,\qquad\text{where}\quad|\Psi\rangle:=\f{\sum_{i=1}^{2^n}|ii\rangle}{\sqrt{2^n}},
    \label{eq: entanglement fidelity def}
\end{equation}
is the so-called entanglement fidelity (equivalent to the purity of the so-called Choi state of $\mc{E}$). Because maximally entangled states can be written in the Pauli basis as $|\Psi\rangle\!\langle\Psi|=4^{-n}\sum_{P\in\mathsf{Pauli}_n}P\otimes{P}^\mathrm{T}$, this is equivalent to
\begin{equation}
    \entfid(\mc{E})=\f{1}{4^n}\sum_{P\in\mathsf{Pauli}_n}\f{1}{2^n}\tr[P\mc{E}(P)] = \f{1}{4^n}\tr(\boldsymbol{\mc{E}}),
    \label{eq: entanglement fidelity and tr PTM}
\end{equation}
where we identified the matrix $\boldsymbol{\mc{E}}$, with elements $\boldsymbol{\mc{E}}_{ij}=\f{1}{2^n}\tr[P_i\mc{E}(P_j)]$, as the Pauli Transfer Matrix representation of $\mc{E}$.

For a depolarizing channel, in particular,
\begin{equation}
    \gamma(\mc{E}_p)=p.
\end{equation}

Also, a composition of depolarizing channels with the same polarization is just another depolarizing channel as
\begin{equation}
    \underbrace{\mc{E}_p\cdots\mc{E}_p}_{k\,\text{times}} = \mc{E}_{p^k}.
\end{equation}

Given the above property, for a depolarizing channel, we also have
\begin{align}
    \unitarity_\psi(\mc{E}_p) &= \f{2^n \gatefid_\psi(\mc{E}_{p^2})-1}{2^n-1} \nonumber\\
    &= \gamma(\mc{E}_p)^2 \nonumber\\
    &:=\unitarity(\mc{E}_p),
    \label{eq: unitarity depolarization}
\end{align}
for any state $|\psi\rangle$. To stress this independence of the state $\psi$, we drop the subindex in the last line, which simply means $\unitarity(\mc{E}_p):=\unitarity_\varphi(\mc{E}_p)$ for an arbitrary pure state $|\varphi\rangle$.

Notice that, $\mc{E}_p\,\mc{U}_\mathsf{C}=\mc{U}_\mathsf{C}\,\mc{E}_p$, or more generally,
\begin{equation}
    \mc{E}_p\,\mc{X}=\mc{X}\mc{E}_p,
    \label{eq: depolarizing commut}
\end{equation}
whenever $\mc{X}$ is a unital and \gls{tp} quantum channel. Furthermore, for any unital and \gls{tp} channel $\mc{X}$, the unitarity $\unitarity_\psi(\mc{E}_p\mc{X})$ only depends on $\gatefid_\psi(\mc{X}^\dg\mc{E}_{p^2}\mc{X})$, which satisfies
\begin{align}
    \gatefid_\psi(\mc{X}^\dg\mc{E}_{p^2}\mc{X}) &= \langle\psi|\mc{X}^\dg\mc{E}_{p^2}\mc{X}(|\psi\rangle\!\langle\psi|)|\psi\rangle \nonumber\\
    &= p^2\gatefid_\psi(\mc{X}^\dg\mc{X}) + (1-p^2)2^{-n}, \nonumber\\
    &= \unitarity(\mc{E}_p)\left(\gatefid_\psi(\mc{X}^\dg\mc{X})-2^{-n}\right)+2^{-n},
\end{align}
and so
\begin{equation}
    \unitarity_\psi(\mc{E}_p\mc{X}) = \f{2^n\unitarity(\mc{E}_p)\left(\gatefid_\psi(\mc{X}^\dg\mc{X})-2^{-n}\right)}{2^n-1}=\unitarity(\mc{E}_p)\unitarity_\psi(\mc{X}),
    \label{eq: unitarity bulk depolarizing}
\end{equation}
i.e. the unitarity this way for depolarizing channels factorizes.

\subsection{Average purity in the many-layer case}\label{appendix: average purity m steps}
We now first consider adding a single layer of gates according to the distribution $\Omega$, and evaluate the purity of $\tilde{\varrho}_1$,
\begin{align}
    \tr(\tilde{\varrho}_1^2) &= \expect_{\substack{\mathsf{L}_1\sim\Omega\\\mathsf{V}\sim\Omega_1}} \tr[\phi(\mathsf{L}_1)\phi(\mathsf{V})(|0\rangle\!\langle0|)\phi(\mathsf{L}_1)\phi(\mathsf{V})(|0\rangle\!\langle0|)^\dg] \nonumber\\
    &= \expect_{\substack{\mathsf{L}_1\sim\Omega\\\psi\sim\Omega_1}}\tr[\mc{E}_1\mc{U}_1\mc{E}_\mathsf{spam}(|\psi\rangle\!\langle\psi|)\mc{E}_1\mc{U}_1\mc{E}_\mathsf{spam}(|\psi\rangle\!\langle\psi|)] \nonumber\\
    &= \expect_{\substack{\mathsf{L}_1\sim\Omega\\\psi\sim\Omega_1}} \gatefid_\psi(\mc{E}_\mathsf{spam}^\dg\mc{U}_1^\dg\mc{E}_1^\dg\mc{E}_1\mc{U}_1\mc{E}_\mathsf{spam}).
\end{align}

This can then be generalized directly to the case of $m$ layers as
\begin{equation}
    \tr(\tilde{\varrho}_m^2) = \expect_{\substack{\mathsf{L}_1\sim\Omega\\\psi\sim\Omega_1}} \gatefid_\psi(\mc{E}_\mathsf{spam}^\dg\mc{U}_1^\dg\mc{E}_1^\dg\cdots\mc{U}_m^\dg\mc{E}_m^\dg\mc{E}_m\mc{U}_m\cdots\mc{E}_1\mc{U}_1\mc{E}_\mathsf{spam}),
    \label{eq: appendix purity m steps no meas}
\end{equation}
and thus by means of Eq.~\eqref{eq: unitarity and fid plus tp}, this is equivalent to
\begin{align}
    &\tr(\tilde{\varrho}_m^2)
    = \left(\f{2^n-1}{2^n}\right)\expect_{\substack{\mathsf{L}_i\sim\Omega\\\psi\sim\Omega_1}}\unitarity_\psi(\mc{E}_m\mc{U}_m\cdots\mc{E}_1\mc{U}_1\mc{E}_\mathsf{spam}) + \mathsf{trace\,decreasing\,terms},
    %\nonumber\\  &+ \f{1}{2^n}\expect_{\psi\sim\Omega_1}\left\{\mathsf{S}_\psi(\mc{E}_\mathsf{spam}^\dg\mc{E}_{\mathsf{avg}_\Omega}^{\dg\,m}\mc{E}_{\mathsf{avg}_\Omega}^m\mc{E}_\mathsf{spam}) + \mathsf{S}_\psi(\mc{E}_{\mathsf{avg}_\Omega}^m\mc{E}_\mathsf{spam})\left[\mathsf{S}_\psi(\mc{E}_{\mathsf{avg}_\Omega}^m\mc{E}_\mathsf{spam}) - \tr[\mc{E}_{\mathsf{avg}_\Omega}^m\mc{E}_\mathsf{spam}(\mbb1/2^n)]\right]\right\},
    \label{eq: general purity decay non-unital non-TP}
\end{align}
where the trace decreasing terms are terms in $\mathsf{S}$ that can be read directly from Eq.~\eqref{eq: unitarity and fid plus tp}. One can express this way the most general form of the purity of the average state $\tilde{\varrho}_m$ in circuit depth $m$.

However, we care, of course, for the situation in which the decay is exponential in $m$ with the unitarity as the rate of decay.
\begin{enumerate}
    \item We may consider first that the noise is approximately gate and time independent, $\mc{E}_i\approx\mc{E}$, in which case at most we can get a double exponential in $\unitarity(\mc{E})$ and $\mc{S}(\mc{E})$, as in~\cite{Wallman_2015}. This approximation can be formalized by assuming that the average over time steps and gates is the leading contribution to the actual noise; see e.g.,~\cite{characterizing_Magesan_2012}.
    \item If we consider noise that is only gate-independent but that is unital and \gls{tp}, we have
\begin{align}
    \tr(\tilde{\varrho}_m^2) &= \expect_{\substack{\mathsf{L}_1\sim\Omega\\\psi\sim\Omega_1}} \gatefid_\psi(\mc{E}_\mathsf{spam}^\dg\mc{U}_1^\dg\mc{E}_1^\dg\cdots\mc{U}_m^\dg\mc{E}_m^\dg\mc{E}_m\mc{U}_m\cdots\mc{E}\mc{U}_1\mc{E}_\mathsf{spam}) \nonumber\\
    &= \expect_{\psi\sim\Omega_1} \langle\psi|\expect_{\mathsf{L}_i\sim\Omega}\mc{E}_\mathsf{spam}^\dg\mc{U}_1^\dg\mc{E}_1^\dg\cdots\mc{U}_m^\dg\mc{E}_m^\dg\mc{E}_m\mc{U}_m\cdots\mc{E}_1\mc{U}_1\mc{E}_\mathsf{spam}(|\psi\rangle\!\langle\psi|)|\psi\rangle,
\end{align}
and we can average recursively over each layer.
    \item\label{appendix: purity to exponential} This circuit averaging, however, becomes significant if the layer set at least forms an $\epsilon$-approximate unitary 2-design~\footnote{ A unitary $t$-design is a probability measure $\mu$ on the $d$-dimensional unitary group $\mbb{U}(d)$ (or a subset thereof), such that $\mc{T}^{(t)}_\mu = \mc{T}_\mathsf{Haar}^{(t)}$, where $\mc{T}^{(t)}_\mu(\cdot):=\int_{\mbb{U}(d)}d\mu(U)U^{\otimes{t}}(\cdot)(U^{\otimes{t}})^\dg$ is a $t$-twirl over the unitary group. There are several nonequivalent, albeit often related, ways of approximating unitary designs, here we employ that in~\cite{Brandao2016, Haferkamp2022randomquantum}, where $(1-\epsilon)\mc{T}_\mu^{(t)}\preccurlyeq\mc{T}^{(t)}_\mathsf{Haar}\preccurlyeq(1+\epsilon)\mc{T}^{(t)}_\mu$ with $A\preccurlyeq{B}$ here if and only if $A-B$ is \gls{cp}, which can be interpreted in a straightforward way as having a small multiplicative factor of difference when measuring a state acted on with $\mc{T}^{(t)}_\mu$ vs with $\mc{T}^{(t)}_\mathsf{Haar}$. Finally, for the case $t=2$, the 2-twirl on a \gls{cp} map $\Phi$ can be stated in terms of its matrix representation $\boldsymbol{\Phi}$ through $\mc{T}_\mu^{(2)}(\boldsymbol{\Phi})$, equivalent to the expression $\int_{\mbb{U}(d)}d\mu(U)U^\dg\Phi(U\rho{U}^\dg)U$ for any state $\rho$.}, to the effect that $\expect_{\mathsf{L}\sim\Omega}\mc{U}^\dg_\mathsf{L}\mc{E}^\dg\mc{E}\mc{U}_\mathsf{L}\approx\tilde{\Xi}_{p}$ for a global depolarizing channel $\tilde{\Xi}_{p^2}$ with polarization
    \begin{align}
        p^2 &= \f{\sum|\tr[K_\alpha^\dg{K}_\beta]|^2-1}{4^n-1}\nonumber\\ &=\f{\tr[\boldsymbol{\mc{E}}^\dg\boldsymbol{\mc{E}}]-1}{4^n-1} \nonumber\\
        &= \f{4^n\entfid(\mc{E}^\dg\mc{E})-1}{4^n-1} \nonumber\\
        &= \f{2^n\overline{\gatefid}(\mc{E}^\dg\mc{E})-1}{2^n-1},
    \end{align}
    where in the first line $\{K_\mu\}$ are the Kraus operators of $\mc{E}$~\cite{Emerson_2005}, or equivalently in the second line with $\boldsymbol{\mc{E}}$ being a matrix representation of $\mc{E}$~\footnote{ It is possible to identify $\boldsymbol{\mc{X}}=\begin{pmatrix} S(\mc{E}) & \boldsymbol{\mc{X}}_{\mathsf{sdl}} \\ \boldsymbol{\mc{X}}_\mathsf{n} & \boldsymbol{\mc{X}}_\mathsf{unital} \end{pmatrix}$, where $S(\mc{E})$, $\boldsymbol{\mc{X}}_{\mathsf{sdl}}$, $\boldsymbol{\mc{X}}_\mathsf{n}$ and $\boldsymbol{\mc{X}}_\mathsf{unital}$ are trace-decreasing, state-dependent leakage, non-unital and unital components (of dimensions 1, $1\times(4^n-1)$, $(4^n-1)\times1$ and square $(4^n-1)$), respectively. This leads to the expression in~\cite{Wallman_2015} for the average unitarity.}, or in terms of the entanglement fidelity through Eq.~\eqref{eq: entanglement fidelity and tr PTM} in the third line, or finally through the average gate-fidelity~\cite{Nielsen_fidelity2002}, as we write in the main text. Thus, if we further consider $\Omega$-distributed circuits that form an approximate unitary 2-design, we may write
    \begin{align}
    \tr(\tilde{\varrho}_m^2) &\approx \expect_{\psi\sim\Omega_1} \langle\psi|\mc{E}_\mathsf{spam}^\dg\tilde{\Xi}_{p_m^2}\cdots\tilde{\Xi}_{p_1^2}\mc{E}_\mathsf{spam}(|\psi\rangle\!\langle\psi|)|\psi\rangle,
\end{align}
where the approximation ``$\approx$'' here neglects the small $\epsilon$ multiplicative corrections; due to the property in Eq.~\eqref{eq: depolarizing commut}, and where we intentionally defined the polarizations to be squares, given that $p^2=\unitarity(\tilde{\Xi}_{p_i})$~\footnote{ We notice that this does not mean that $\tilde{\Xi}_{p}$ corresponds to (approximately) the average of a single $\mc{E}$ (which only occurs if $\mc{E}$ itself is already depolarizing).}. For approximately time-stationary noise, to the effect that all polarizations are equal, $p_1=\cdots=p_m=p$, this is equivalent to the exponential decay
\begin{align}
    \tr(\tilde{\varrho}_m^2) &\approx A\unitarity(\tilde{\Xi}_p)^m+\f{1}{2^n},
\end{align}
where $A=2^{-n}(2^n-1)\expect_{\psi\sim\Omega_1}\unitarity_\psi(\mc{E}_\mathsf{spam})$ is a \gls{spam} constant. Of course, given that $\unitarity(\tilde{\Xi}_p)=p^2$, this unitarity can also be expressed by $\unitarity(\mc{E})$, as done in Eq.~\eqref{eq: main exponential decay} in the main text.
\end{enumerate}

\section{Operational estimation of the purity}\label{appendix: operational estimation of the purity}
Now we know how the purity of the noisy average state $\tilde{\varrho}_m$ in Eq.~\eqref{eq: avg state layer m} behaves in terms of the average unitarity of noise with respect to circuit depth $m$. The task is then to estimate this quantity via \gls{rm}s.

\subsection{Purity estimation through randomized measurements}
\gls{rm}s found some of their first applications in estimating the purity of the quantum state of a subsystem~\cite{vanEnk_2012, Renyi_2018}. Specifically, here we will focus on the results of~\cite{purity_science_2019, purity_pra_2019}, whereby it is shown that for a reduced quantum state $\rho_A=\tr_B(\rho_{AB})$, from some composite closed quantum system $AB$, where here we assume $A$ to be a $n$-qubit quantum system, its purity can be expressed as
\begin{equation}
    \tr(\rho_\mathsf{A}^2) = 2^{n}\sum_{\mathbf{s},\mathbf{{s}^\prime}} (-2)^{-h(\mathbf{s},\mathbf{s}^\prime)}\expect_{\mathsf{U}\sim\mathsf{Haar}}\left[\mathsf{P}_\mathsf{U}(\mathbf{s})\mathsf{P}_\mathsf{U}(\mathbf{s}^\prime)\right],
    \label{eq: appendix purity random cross-correlations}
\end{equation}
where $\mathbf{s}$ and $\mathbf{s}^\prime$ are $n$-bit strings, $h(\mathbf{s},\mathbf{s}^\prime)$ is the Hamming distance (number of distinct bits) between them, $\expect_{\mathsf{U}\sim\mathsf{Haar}}$ denotes uniform averaging with the Haar measure over the local unitaries $\mathsf{U}=\Motimes_{i=1}^n\mathsf{U}_i$,~\footnote{ That is, $\mbb{E}_\mathsf{Haar}$ here refers to a uniform and independent average over all the local unitaries $U_i$ with the Haar measure, i.e., $\mbb{E}_\mathsf{Haar}\sim\mbb{E}_{U_1}\mbb{E}_{U_2}\cdots\mbb{E}_{U_{N}}$} and the
\begin{equation}
\mathsf{P}_\mathsf{U}(\mathbf{s}):=\langle\mathbf{s}|\,\mc{U}_\mathsf{U}(\rho_\mathsf{A})|\mathbf{s}\rangle,
    \label{eq: appendix probability purity orig}
\end{equation}
are noiseless probabilities of observing the $n$-bit string $\mathbf{s}$ upon applying the unitary $\mathsf{U}$ on the reduced state $\rho_\mathsf{A}$.

The main advantage here is that the quantum component of estimating purity this way is just the probabilities in Eq.~\eqref{eq: appendix probability purity orig}, which simply means applying a product of random unitaries and measuring each on a user-specified basis. The rest, i.e., fully estimating Eq.~\eqref{eq: appendix purity random cross-correlations} is purely classical post-processing of said probabilities. For $n$ qubits, the number of distinct terms in the sum over $n$-bit strings in Eq.~\eqref{eq: appendix purity random cross-correlations} is $2^{n-1}(2^n-1)$, since the Hamming distance is symmetric, $h(\mathbf{s},\mathbf{s}^\prime) = h(\mathbf{s}^\prime, \mathbf{s})$; however, this sum is only a function of the probabilities $\mathsf{P}_\mathsf{U}$, so it could, for example, be computed in parallel.

\subsection{Purity of many layer \texorpdfstring{$\Omega$}{Omega}-distributed circuits through randomized measurements}\label{appendix: purity with rms}
Let us again begin by considering the case $m=1$, of a single random layer, so that the equivalent of the probability in Eq.~\eqref{eq: appendix probability purity orig}, by measuring the state $\tilde{\varrho}_1$ with a \gls{rm} using a random unitary $\mathsf{W}\in\mbb{L}(\mbb{G}_1)$ over $\Omega_1$, reads as
\begin{align}
    \mathsf{P}_{1,\mathsf{W}}(\mathbf{s}) &= \tr[\phi(\mathsf{W})(|\mathbf{s}\rangle\!\langle\mathbf{s}|)\,\tilde{\varrho}_1] \nonumber\\
    &= \expect_{\substack{\mathsf{L}_1\sim\Omega\\\mathsf{V}\sim\Omega_1}} \tr[\phi(\mathsf{W})(|\mathbf{s}\rangle\!\langle\mathbf{s}|)\,\phi(\mathsf{L}_1)\phi(\mathsf{V})(|0\rangle\!\langle0|)] \nonumber\\    &=\expect_{\substack{\mathsf{L}_1\sim\Omega\\\psi\sim\Omega_1}} \tr[\mc{E}_{\mathsf{spam}_\mathsf{w}}^\prime\mc{U}^\prime_\mathsf{W}(|\mathbf{s}\rangle\!\langle\mathbf{s}|)\,\mc{E}_1\,\mc{U}_1\,\mc{E}_{\mathsf{spam}_\mathsf{v}}(|\psi\rangle\!\langle\psi|)] \nonumber\\ &=\expect_{\substack{\mathsf{L}_1\sim\Omega\\\psi\sim\Omega_1}} \langle\mathbf{s}|\,\mc{U}_\mathsf{W}\,\mc{E}_{\mathsf{spam}_\mathsf{w}}\,\mc{E}_1\,\mc{U}_1\,\mc{E}_{\mathsf{spam}_\mathsf{v}}(|\psi\rangle\!\langle\psi|)|\mathbf{s}\rangle,
\end{align}
where we again defined $\mc{U}_{\mathsf{V}}(|0\rangle\!\langle0|)=|\psi\rangle\!\langle\psi|$ as a random (according to a uniform $\mc{U}_{\mathsf{V}}\sim\Omega_1$) pure state, and identified each $\mc{E}_\mathsf{spam}$ as the noise contribution to \gls{spam}; in the penultimate line (purely for notational reasons) we defined $\phi(\mathsf{W}):=\mc{E}_{\mathsf{spam}_\mathsf{w}}^\prime\mc{U}^\prime_\mathsf{W}$ and in the last line, $\mc{U}_\mathsf{W}=\mc{U}_\mathsf{W}^{\prime\,\dg}$ and $\mc{E}_{\mathsf{spam}_\mathsf{w}}=\mc{E}_{\mathsf{spam}_\mathsf{w}}^{\prime\,\dg}$.

Putting this together with the probability of getting another $n$-bit string $\mathbf{s}^\prime$ with the same layer sequence,
\begin{align}
    \mathsf{P}_{1,\mathsf{C}}(\mathbf{s})\mathsf{P}_{1,\mathsf{C}}(\mathbf{s}^\prime)
    &= \expect_{\substack{\mathsf{L}_1\sim\Omega\\\psi\sim\Omega_1}} \langle\mathbf{s}|\mc{U}_\mathsf{W}\,\mc{E}_{\mathsf{spam}_\mathsf{w}}\,\mc{E}_1\,\mc{U}_1\,\mc{E}_{\mathsf{spam}_\mathsf{v}}(|\psi\rangle\!\langle\psi|)|\mathbf{s}\rangle\!\langle\mathbf{s}^\prime|\,\mc{U}_\mathsf{W}\,\mc{E}_{\mathsf{spam}_\mathsf{w}}\,\mc{E}_1\,\mc{U}_1\,\mc{E}_{\mathsf{spam}_\mathsf{v}}(|\psi\rangle\!\langle\psi|)|\mathbf{s}^\prime\rangle \nonumber\\
    &= \expect_{\substack{\mathsf{L}_1\sim\Omega\\\psi\sim\Omega_1}}\langle\mathbf{s}|\,\mc{U}_\mathsf{W}\,\mc{E}_{\mathsf{spam}_\mathsf{w}}\,\mc{E}_1\,\mc{U}_1\,\mc{E}_{\mathsf{spam}_\mathsf{v}}[|\psi\rangle\!\langle\psi|\mc{E}_{\mathsf{spam}_\mathsf{v}}^\dg\,\mc{U}_1^\dg\mc{E}_1^\dg\,\mc{E}_{\mathsf{spam}_\mathsf{w}}^\dg\,\mc{U}_\mathsf{W}^\dg(|\mathbf{s}^\prime\rangle\!\langle\mathbf{s}^\prime|)|\psi\rangle\!\langle\psi|]|\mathbf{s}\rangle\nonumber\\
    &= \expect_{\substack{\mathsf{L}_1\sim\Omega\\\psi\sim\Omega_1}}\langle\mathbf{s}|\,\mc{U}_\mathsf{W}\,\mc{E}_{\mathsf{spam}_\mathsf{w}}\,\mc{E}_1\,\mc{U}_1\,\mc{E}_{\mathsf{spam}_\mathsf{v}}\mc{P}_\psi\mc{E}_{\mathsf{spam}_\mathsf{v}}^\dg\,\mc{U}_1^\dg\mc{E}_1^\dg\,\mc{E}_{\mathsf{spam}_\mathsf{w}}^\dg\,\mc{U}_\mathsf{W}^\dg(|\mathbf{s}^\prime\rangle\!\langle\mathbf{s}^\prime|)|\mathbf{s}\rangle,
    \label{eq: probability correlations m=1}
\end{align}
where in the last line we defined $\mc{P}_\psi(\cdot):=|\psi\rangle\!\langle\psi|\cdot|\psi\rangle\!\langle\psi|$.

Now we can exploit the identity derived in~\cite{proctor2022establishing},
\begin{align}
    \entfid(\mc{X})&= \sum_x\left(-2\right)^{-h(x,y)}\langle{x}|\expect_{\mathsf{U}_i\sim\mathsf{Haar}}\left\{\mc{U}_\mathsf{U}^\dg\,\mc{X}\,\mc{U}_\mathsf{U}\right\}\,[|y\rangle\!\langle{y}]|x\rangle,
    \label{eq: average product 2-designs}
\end{align}
where here $\entfid$ is entanglement fidelity, defined in Eq.~\eqref{eq: entanglement fidelity def}, and $\mc{U}_\mathsf{U}$ is the noiseless unitary map corresponding to the product $\mathsf{U}=\Motimes_{i=1}^n\mathsf{U}_i$ with local, single-qubit random unitaries $\mathsf{U}_i$. Since this expression requires two copies of $\mc{U}_\mathsf{U}$, where only the individual $\mathsf{U}_i$ should be Haar random, it suffices for these (as opposed to the global unitary) to belong to a unitary 2-design, e.g., the single-qubit Clifford group.

We emphasize then, that we will require the \gls{rm}s to be randomized by a layer made up of uniformly random single-qubit Clifford gates; while in the main text we fix this to correspond to $\mbb{G}_1$ and $\Omega_1$, all other layers in principle can be chosen arbitrarily (with the reasoning of having some better choices than others detailed in the previous sections).

We can now apply Eq.~\eqref{eq: average product 2-designs} on Eq.~\eqref{eq: probability correlations m=1}, so that
\begin{align}
    \tr(\tilde{\varrho}_1^2) &\simeq 4^n \expect_{\substack{\mathsf{L}_1\sim\Omega\\\psi\sim\Omega_1}}\entfid(\mc{E}_{\mathsf{spam}_\mathsf{w}}\mc{E}_1\mc{U}_1\mc{E}_{\mathsf{spam}_\mathsf{v}}\mc{P}_\psi\mc{E}_{\mathsf{spam}_\mathsf{v}}^\dg\,\mc{U}_1^\dg\mc{E}_1^\dg\mc{E}_{\mathsf{spam}_\mathsf{w}}^\dg) \nonumber\\
    &= \expect_{\substack{\mathsf{L}_1\sim\Omega\\\psi\sim\Omega_1}}\gatefid_\psi(\mc{E}_{\mathsf{spam}_\mathsf{v}}^\dg\,\mc{U}_1^\dg\mc{E}_1^\dg\mc{E}_{\mathsf{spam}_\mathsf{w}}^\dg\mc{E}_{\mathsf{spam}_\mathsf{w}}\mc{E}_1\,\mc{U}_1\mc{E}_{\mathsf{spam}_\mathsf{v}}),
    \label{eq: avg purity 1-step}
\end{align}
where the second line follows directly from the definition of entanglement fidelity. This generalizes to any circuit depth $m$, to
\begin{equation}
    \tr(\tilde{\varrho}_m^2) = \expect_{\substack{\mathsf{L}_i\sim\Omega\\\psi\sim\Omega_1}}\gatefid_\psi(\mc{E}_{\mathsf{spam}_\mathsf{v}}^\dg\,\mc{U}_1^\dg\mc{E}_1^\dg\cdots\mc{U}_m^\dg\mc{E}_m^\dg\mc{E}_{\mathsf{spam}_\mathsf{w}}^\dg\mc{E}_{\mathsf{spam}_\mathsf{w}}\mc{E}_m\,\mc{U}_m\cdots\mc{E}_1\,\mc{U}_1\mc{E}_{\mathsf{spam}_\mathsf{v}}),
    \label{eq: purity in fid}
\end{equation}
which coincides with Eq.~\eqref{eq: appendix purity m steps no meas} up to $\mc{E}_{\mathsf{spam}_\mathsf{w}}$.

Now similar to point~\ref{appendix: purity to exponential} in \S~\ref{appendix: average purity m steps} above, we mainly care about an exponential decay, which occurs for $\Omega$-distributed circuits being an approximate unitary 2-design and noise being approximately gate+time independent, unital and \gls{tp}. Now we will have
\begin{equation}
    \tr(\tilde{\varrho}_m^2) \approx A\unitarity(\tilde{\Xi}_p)^m + \f{1}{2^n},
    \label{eq: appendix exponential decay}
\end{equation}
for $A=2^{-n}(2^n-1)\unitarity(\tilde{\Xi}_{q_{\mathsf{spam}_w}})\mbb{E}_{\psi\sim\Omega_1}\unitarity_\psi(\mc{E}_{\mathsf{spam}_\mathsf{v}})$, where $\tilde{\Xi}_{q_{\mathsf{spam}_\mathsf{w}}}$ is the contribution from the layer average of $\mc{E}_{\mathsf{spam}_\mathsf{w}}$; of course both unitarity terms in $A$ cannot be distinguished and can simply be interpreted as an average unitarity of \gls{spam}.

The result in Eq.~\eqref{eq: appendix exponential decay} corresponds to that with an exact average over layers and initial states; this average can be approximated numerically with Eq.~\eqref{eq: main average purity estimation in depth} given a number $N_\mathsf{W}$ of measurements, $\mathsf{C}_m$ of $\Omega$-distributed circuits, and $N_\mathsf{V}$ of initial randomization samples, respectively.

% From here, it is clear that
% \begin{align}
%     \tr\left[\phi(\tilde{\sigma_m})^2\right] &\simeq \expect_{|\psi\rangle} \gatefid_\psi(\mc{E}_\mathsf{spam}\,\mc{E}^{\dg\,m}_{\mathsf{avg}_\Omega}\mc{E}_\mathsf{spam}^2\mc{E}^m_{\mathsf{avg}_\Omega}\,\mc{E}_\mathsf{spam}).
% \end{align}

% Now, the generalization of Eq.~\eqref{eq: purity m general spam} might take a more complicated form in the trace-decreasing part, however, we are really interested in generalizing the effectively \gls{tp} case in Eq.~\eqref{eq: purity m TP}, for which we simply get
% \begin{align}
%     \tr\left[\phi(\tilde{\sigma}_m)^2\right] \simeq \left(\f{2^N-1}{2^N}\right)\gamma(\mc{E}_\mathsf{spam})^4\expect_{|\psi\rangle}u_\psi(\mc{E}_{\mathsf{avg}_\Omega}^m) + \f{1}{2^N},
% \end{align}
% i.e., under this noise model the \gls{spam} errors would at most affect the unitarity decay with a multiplicative constant.

\section{Operational estimation of fidelity through randomized measurements}\label{appendix: operational estimation of fidelity}
Since estimating the average unitarity is really useful only when informed by the average fidelity, we now consider estimating the average layer fidelity of the same $\Omega$-distributed circuits, once given a set of randomized measurement probabilities $\mathsf{P}_\mathsf{U}$.

Naturally, if the noiseless output is a state $|\varphi\rangle$,  Eq.~\eqref{eq: appendix purity random cross-correlations} can be modified to account for its fidelity with respect to a mixed state $\rho$ as
\begin{equation}
    \langle\varphi|\rho|\varphi\rangle = 2^{n}\sum_{\mathbf{s},\mathbf{{s}_\mathsf{ideal}}} (-2)^{-h(\mathbf{s},\mathbf{s}_\mathsf{ideal})}\expect_{\mathsf{U}\sim\mathsf{Haar}}\mathsf{P}_\mathsf{U}(\mathbf{s})\mathsf{P}^{(\mathsf{ideal})}_\mathsf{U}(\mathbf{s}_\mathsf{ideal}),
    \label{eq: appendix fidelity random measurements}
\end{equation}
where $\mathsf{P}^{(\mathsf{ideal})}_\mathsf{U}$ is the probability of observing a $n$-bit string $\mathbf{s}_\mathsf{ideal}$ with a randomized measurement defined by $\mathsf{U}$ on $|\varphi\rangle\!\langle\varphi|$, that is
\begin{equation}
    \mathsf{P}^{(\mathsf{ideal})}_\mathsf{U}(\mathbf{s}_\mathsf{ideal}) = |\langle\mathbf{s}_\mathsf{ideal}|\mathsf{U}|\varphi\rangle|^2.
\end{equation}

Notice that this assumes that we can efficiently compute the $n$-bit strings, and thus the probabilities, corresponding to the noiseless outputs. This generally implies a limitation in the gate sets that can be employed to only Clifford gates, however, for mid-scale estimations the ideal probabilities can still be computed without a major overhead when the gate set contains non-Clifford gates to be sampled with a relatively low probability.

In our case, we want to estimate the quantity $\langle\psi_m|\tilde{\varrho}_m|\psi_m\rangle$, where
\begin{align}
    |\psi_m\rangle\!\langle\psi_m| := \mathsf{C}_m^{(\mathsf{ideal})}(|0\rangle\!\langle0|) &= \mc{U}_m\cdots\mc{U}_1\mc{U}_\mathsf{V}(|0\rangle\!\langle0|) \nonumber\\
    &= \mc{U}_m\cdots\mc{U}_1(|\psi\rangle\!\langle\psi|).
\end{align}

Thus, consider first the case $m=1$; we have
\begin{align}
    \mathsf{P}_{1,\mathsf{W}}(\mathbf{s})\mathsf{P}^{(\mathsf{ideal})}_{1,\mathsf{W}}(\mathbf{s}_\mathsf{ideal}) &= \expect_{\substack{\mathsf{L}_1\sim\Omega\\\psi\sim\Omega_1}}\langle\mathbf{s}|\,\mc{U}_\mathsf{W}\,\mc{E}_{\mathsf{spam}_\mathsf{w}}\,\mc{E}_1\,\mc{U}_1\,\mc{E}_{\mathsf{spam}_\mathsf{v}}\mc{P}_\psi\,\mc{U}_1^\dg\,\mc{U}_\mathsf{W}^\dg(|\mathbf{s}_\mathsf{ideal}\rangle\!\langle\mathbf{s}_\mathsf{ideal}|)|\mathbf{s}\rangle,
\end{align}
which follows as in Eq.~\eqref{eq: probability correlations m=1}, where in the last line we defined $\mc{P}_\psi(\cdot):=|\psi\rangle\!\langle\psi|\cdot|\psi\rangle\!\langle\psi|$. Similarly, we can follow all steps from Eq.~\eqref{eq: probability correlations m=1} to Eq.~\eqref{eq: purity in fid} in an analogous way to obtain
\begin{equation}
    \langle\psi_m|\tilde{\varrho}_m|\psi_m\rangle = \expect_{\substack{\mathsf{L}_i\sim\Omega\\\psi\sim\Omega_1}}\gatefid_\psi(\,\mc{U}_1^\dg\cdots\mc{U}_m^\dg\mc{E}_{\mathsf{spam}_\mathsf{w}}\mc{E}_m\,\mc{U}_m\cdots\mc{E}_1\,\mc{U}_1\mc{E}_{\mathsf{spam}_\mathsf{v}}),
\end{equation}
which is simply an \gls{rb}-like fidelity decay. Similarly here, if the $\Omega$-distributed circuits approximate a unitary 2-design, to the effect that $\expect_{\mathsf{L}\sim\Omega}\mc{U}^\dg_\mathsf{L}\mc{E}_i\mc{U}_\mathsf{L}\approx\tilde{\Xi}_{p_i}$, i.e., the average noise is approximately a global depolarizing channel with $p_i=(2^n\overline{\gatefid}(\mc{E})-1)/(2^n-1)$, we end up with
\begin{align}
\langle\psi_m|\tilde{\varrho}_m|\psi_m\rangle &\approx \expect_{\psi\sim\Omega_1}\gatefid_\psi(\tilde{\Xi}_{p_m^\prime}\tilde{\Xi}_{p_{m-1}}\cdots\tilde{\Xi}_{p_1}\mc{E}_{\mathsf{spam}_\mathsf{v}}) \nonumber\\
&= \expect_{\psi\sim\Omega_1}\gatefid_\psi(\tilde{\Xi}_{p_{\mathsf{spam}_\mathsf{w}}}\tilde{\Xi}_{p_m{p}_{m-1}\cdots{p}_1}\mc{E}_{\mathsf{spam}_\mathsf{v}}),
\end{align}
 which is a \gls{rb}-like decay of the state fidelity in the circuit depth $m$, where we defined $p_m^\prime=p_mp_{\mathsf{spam}_\mathsf{w}}$ for some $p_{\mathsf{spam}_\mathsf{w}}$. Further, for time-independent noise, such that $p_1=p_2=\ldots=p_m=p$, this is an exponential decay in $m$,
\begin{align}
\langle\psi_m|\tilde{\varrho}_m|\psi_m\rangle &\approx Ap^m + B,
\end{align}
where $A=p_{\mathsf{spam}_\mathsf{w}}[\gatefid_\psi(\mc{E}_{\mathsf{spam}_\mathsf{v}})-2^{-n}]$, $B=2^{-n}$, and
\begin{equation}
    p = \f{2^n\overline{\gatefid}(\mc{E})-1}{2^n-1},
\end{equation}
which can similarly be written in terms of entanglement fidelity of $\mc{E}$, or in terms of its matrix representation, similar to the usual \gls{rb} constants for unital \gls{spam}.

This means we can generate a set of random $\Omega$-distributed circuits, attach some set of randomized measurements, and use the probabilities to simultaneously estimate average layer fidelity and unitarity.

\section{Scrambling and unitary 2-designs}\label{appendix: unitary designs}
%\label{sec: sample complexity and unitary designs}
Here we refer to both assumptions, \ref{assumption: tp unital},~\ref{assumption: 2-design}, as written in the main text in \S~\ref{sec: conditions for exponential}, for an exponential decay of the average sequence purity in circuit depth.

While assumption~\ref{assumption: tp unital} is beyond the choices that the user executing the protocol~\ref{sec: protocol} can make, assumption~\ref{assumption: 2-design} is important to make the protocol~\ref{protocol: classical post-processing} adhere to the exponential decay of main Result~\ref{result: exponential decay}, but it also sets a practical limit in scalability to mid-scale systems, which we discuss below.

There is, furthermore, a fundamental limit to scalability imposed by the \emph{sample complexity} of estimating purity via \gls{rm}s, i.e., the number of experimental samples, $N_\mathsf{meas}$ in Eq.~\eqref{eq: probability estimator}, needed to estimate purity within a given error. This is discussed in the main text in \S~\ref{sec: sample complexity}, and it compounds with the sample complexity of the $\Omega$-distributed circuits, related to assumption~\ref{assumption: 2-design}.

The need for unitary 2-designs in \gls{rb}-based techniques stems from the fact that, by definition, these inherit the property of fully random (i.e. Haar distributed) unitaries whereby their second moment is described by a depolarizing channel, which requires a single parameter to be fully specified. This implies that figures of merit, such as average gate fidelity or average gate unitarity can be encapsulated into such polarization parameter (once assumption~\ref{assumption: tp unital} is satisfied) given circuits that generate a unitary 2-design.

Loosely speaking, a unitary 2-design is a probability distribution $\mu$ on the unitary group, or a subset thereof, satisfying $\expect_{V\sim\mu}\mc{V}^\dg\mc{X}\mc{V} = \expect_{U\sim\mathsf{Haar}}\mc{U}^\dg\mc{X}\mc{U}$ for any quantum channel $\mc{X}$. The action $\expect_{V\sim\mu}\mc{V}^\dg(\cdot)\mc{V}$ is referred to as a 2-twirl with $\mu$. That is, the 2-twirl with $\mu$ reproduces the second moment of the whole unitary group with the Haar measure. That an ensemble of unitaries constitutes a 2-design is particularly useful because it is known that doing a 2-twirl with the Haar measure on a \gls{cp} map, reduces it to a depolarizing channel~\cite{Emerson_2005}, i.e., also for a 2-design $\expect_{V\sim\mu}\mc{V}^\dg\mc{X}\mc{V}(\cdot) = p (\cdot)+(1-p)\mbb1/2^n$, where here $p=(2^n\overline{\gatefid}(\mc{X})-1)/(2^n-1)$, where $\overline{\gatefid}(\mc{X})$ is the average gate fidelity of $\mc{X}$. Both Eq.~\eqref{eq: sq noise average} and Eq.~\eqref{eq: average polarization} are obtained using this relation.

The notion of a unitary 2-design can further be relaxed by having equality up to a small $\epsilon$ according to a given metric, to the effect that the 2-twirl is also \emph{close} to a depolarizing channel. More precisely, denoting a 2-twirl by $\Delta_\mu(\cdot):=\expect_{V\sim\mu}\mc{V}^\dg(\cdot)\mc{V}$ we say that $\mbb{V}$ is an $\epsilon$-approximate unitary 2-design, if
\begin{equation}
    (1-\epsilon)\Delta_\mathsf{Haar}\preccurlyeq\Delta_\mu\preccurlyeq(1+\epsilon)\Delta_\mathsf{Haar},
\end{equation}
where here $\preccurlyeq$ is semidefinite ordering in the sense that $\mc{X}\preccurlyeq\mc{Y}$ if and only if $\mc{X}-\mc{Y}$ is a \gls{cp} map. This definition was put forth (for the more general setting of $t$-designs) in~\cite{Brandao2016} and has the particularly simple interpretation of having a small multiplicative factor $1\pm\epsilon$ of difference when measuring a state acted on with a channel twirled $\Delta_\mu$ as opposed to with $\Delta_\mathsf{Haar}$~\footnote{ This definition is further connected in~\cite{Brandao2016} to the usual diamond norm definition as implying that if $\mbb{V}$ is an $\epsilon$-approximate unitary 2-design, also $\|\Delta_\mu-\Delta_\mathsf{Haar}\|_\diamond\leq2\epsilon$, where $\|\mc{X}\|_\diamond:=\sup_\rho\|(\mc{I}_d\otimes\mc{X})\rho\|_1$, where $\|X\|_1=\tr\sqrt{XX^\dg}$ denotes trace norm and the supremum is taken over all dimensions $d\geq1$ of the identity and corresponding density matrices $\rho$.}. Thus, the approximation sign in both Eq.~\eqref{eq: main exponential decay} and Eq.~\eqref{eq: fidelity decay} refers to these small multiplicative factors in the 2-design approximation, exponential in $m$.

A well-known example of an exact 2-design is the $n$-qubit Clifford group~\cite{PhysRevA.80.012304}: it is also simultaneously a strong reason
why standard~\footnote{ As can be seen in~\cite{helsen_general} and~\cite{FigueroaRomero2022towardsgeneral}, there is a plethora of techniques under the term \gls{rb}; by standard we mean single or two-qubit RB with corresponding Clifford gate sets, estimating the group's average gate fidelity by fitting survival probabilities to an exponential decay in sequence length.} Clifford \gls{rb} works so well, and the culprit (in practice) of why it does not scale in $n$. Techniques akin to standard Clifford \gls{rb}, such as \gls{drb}, \gls{mrb} and \gls{birb} resolve such scalability issues by, among other things, retaining some of the randomness through $\Omega$-distributed circuits. Rather than requiring an approximate unitary design property, e.g., \gls{mrb}~\cite{hines2022demonstrating} and \gls{birb}~\cite{hines2023fully} require a scrambling property in the sense that Pauli errors get quickly spread among other qubits before other Pauli errors occur. This can be stated in terms of an entanglement fidelity for non-identity Paulis $P$, $P^\prime$, with action $\mc{P}(\cdot)=P(\cdot)P$ and $\mc{P}\prime(\cdot)=P^\prime(\cdot)P^\prime$, such that for an expected infidelity of the layers $\alpha$, there is $k\ll1/\alpha$ and a $\delta\ll1$ with
\begin{equation}
\expect_{\Omega}\entfid\!\left[\mc{P}\,\mc{C}_\Omega\,\mc{P}^\prime\,\mc{C}_\Omega^{-1}\right] \leq \delta + \f{1}{4^n},
\label{eq: scrambling cond}
\end{equation}
where here $\mc{C}_\Omega=\mc{U}(\mathsf{L}_k\cdots\mathsf{L}_1)$, with $\mc{U}$ the unitary map of the sequence of noiseless layers $\mathsf{L}_k\cdots\mathsf{L}_1$. Similarly, \gls{drb} uses \emph{generators} of unitary 2-designs as the benchmarking gate set and then requires that these constitute a \emph{sequence-asymptotic} unitary 2-design~\cite{directRB2023}.

While the scrambling condition in Eq.~\eqref{eq: scrambling cond} is stated for Pauli noise channels, the choice of $(\mbb{G}_1,\Omega_1)$ being the uniformly distributed single-qubit Clifford group has the effect of projecting any \gls{cp} map (modeling Markovian noise) to a Pauli channel (i.e., a tensor product of depolarizing channels) since each Clifford group constitutes a unitary 2-design on the respective qubit.

The reason approximating a unitary 2-design is a stronger condition than scrambling, as defined by Eq.~\eqref{eq: scrambling cond}, is because it would require $\delta$ shrinking as $\mc{O}(4^{-n})$~\footnote{ This can be seen e.g., by assuming the $\mc{U}(\mathsf{L}_k\cdots\mathsf{L}_1)$ generates a unitary 2-design and using Eq.~\eqref{eq: entanglement fidelity and tr PTM}}. Nevertheless, while we directly assume that our $\Omega$-distributed circuits approximate a unitary 2-design, for a mid-scale of tenths of qubits the scrambling condition suffices, and indeed it holds in a similar way than it does in \gls{mrb} or \gls{birb}~\footnote{ A difference of our technique with \gls{mrb} is that we do not employ a mirror structure, while one with \gls{birb} is that we do not employ initial and final stabilizers.}, as exemplified in \S~\ref{sec: demonstration}. Increasing the qubit count to estimate average unitarity would require establishing a scrambling condition, e.g., as for \gls{birb} by employing stabilizer states and measurements, or otherwise relaxing the approximate design condition.

\section{Numerical addendum}\label{appendix: numerical addendum}
\subsection{The simulator backend}\label{appendix: simulator backend}
\begin{figure}[ht!]
     \centering
     \includegraphics[width=0.3\textwidth]{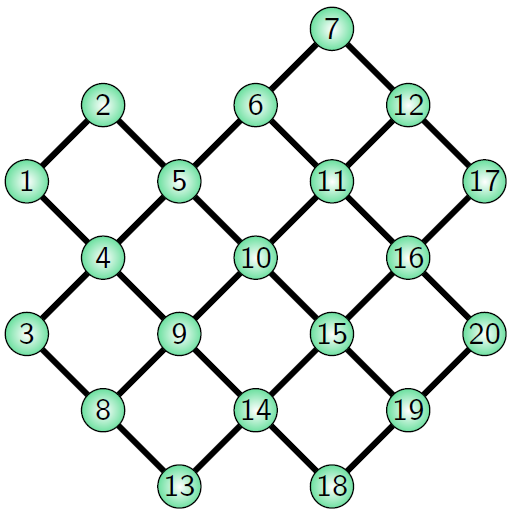}
     \caption{\textbf{Backend topology and connectivity}: We consider a backend with 20 possible qubits (circles), with each qubit labeled with a given integer and connected to another as depicted by the corresponding edges. We employ $\Omega$-distributed circuits with only single- and two-qubit gates, with two-qubits gates acting only between pairs of connected qubits.}
     \label{fig: backend topology}
 \end{figure}

As explained in the main text in \S~\ref{sec: demonstration}, we performed experiments on \spark, a 5-qubit superconducting system with a central qubit connected to the rest, and simulations with a backend of a total of 20 qubits in a grid topology, connected as depicted in the graph of Fig.~\ref{fig: backend topology}. In the simulation, we considered a subset of these qubits, namely qubits ranging from qubit 3 to qubit 12, in pairs as depicted in Fig.~\ref{fig: purity}\textsf{(b)}. All qubits were associated with a noise model on readout, gate times, single- and two-qubit depolarizing noise, and decoherence, $T_1$ and $T_2$ times. 

 \subsection{Simulation on 10 qubits}
\begin{figure*}[ht!]
    \centering \includegraphics[width=\textwidth]{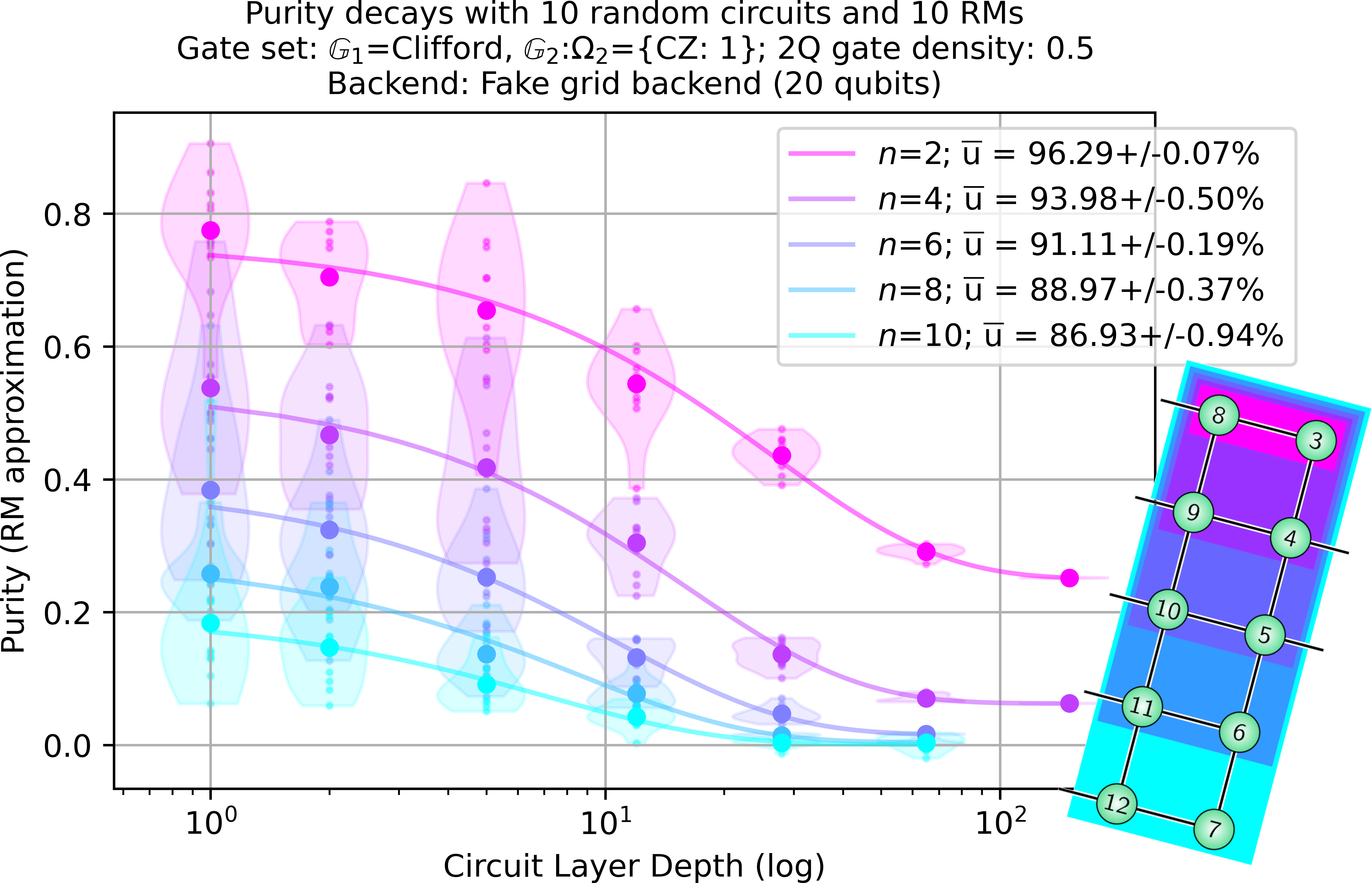}
    \caption{\textbf{Purity decays for simulated backend}. Numerical simulation for purity decays computed according to Protocol~\ref{sec: protocol} for $\Omega$-distributed circuits using the gate set $\mbb{G}=\mathsf{Clif}_1\cup\{\mathsf{CZ}\}$, with $\mathsf{Clif}_1$ being the uniformly-distributed single-qubit Clifford group. Individual plots correspond to a given number of qubits $n$, taken from a connectivity graph shown in (b), whereby smaller points correspond to \gls{moms} estimators of purity with Eq.~\eqref{eq: main average purity estimation in depth} at a given depth $m$, bigger points correspond to the average purity, violins show the distribution of the \gls{moms} estimators, and lines correspond to a least-squares fit of the averages to the decay model in Eq.~\eqref{eq: main exponential decay}, whereby the average unitarities $\overline{\unitarity}$ in corresponding number of qubits $\mathsf{n}$ are estimated. Sample parameters: $N_{\mathsf{C}_m^*}=12$, $N_\mathsf{W}=9$, $N_{\mathsf{meas}}=2^{10}$ and $K=9$ \gls{moms}.}
    \label{fig: purity}
\end{figure*}
We now employ a simulator backend that considers 20 possible qubits with a grid topology, as displayed fully in Fig.~\ref{fig: backend topology} of Appendix~\ref{appendix: simulator backend}. All operations (except barriers) are modeled as noisy, in this particular example with a set of parameters considering a range of relaxation times $T_1$, dephasing times $T_2$, gate duration parameters, depolarizing parameters on both single- and two-qubit gates, and readout errors (bit-flip probabilities).

Similarly in this case, according to Protocol~\ref{sec: protocol}, we extracted decay rates of the average state purity and fidelity in increasing depths, and thus average layer unitarity and fidelity for $\Omega$-distributed circuits with up to 10 qubits. In Fig.~\ref{fig: purity}, we show the corresponding purity decays in increasing depth. We display only even qubit numbers $n$ for easiness of visualization, where each $n$ corresponds to choices of qubits and couplings as shown in the inset. In all cases, we employed unbiased estimators and $K=2$ \glsreset{moms}\gls{moms}\glsreset{moms} estimators.

\begin{figure*}[t!]
    \centering
    \includegraphics[width=\textwidth]{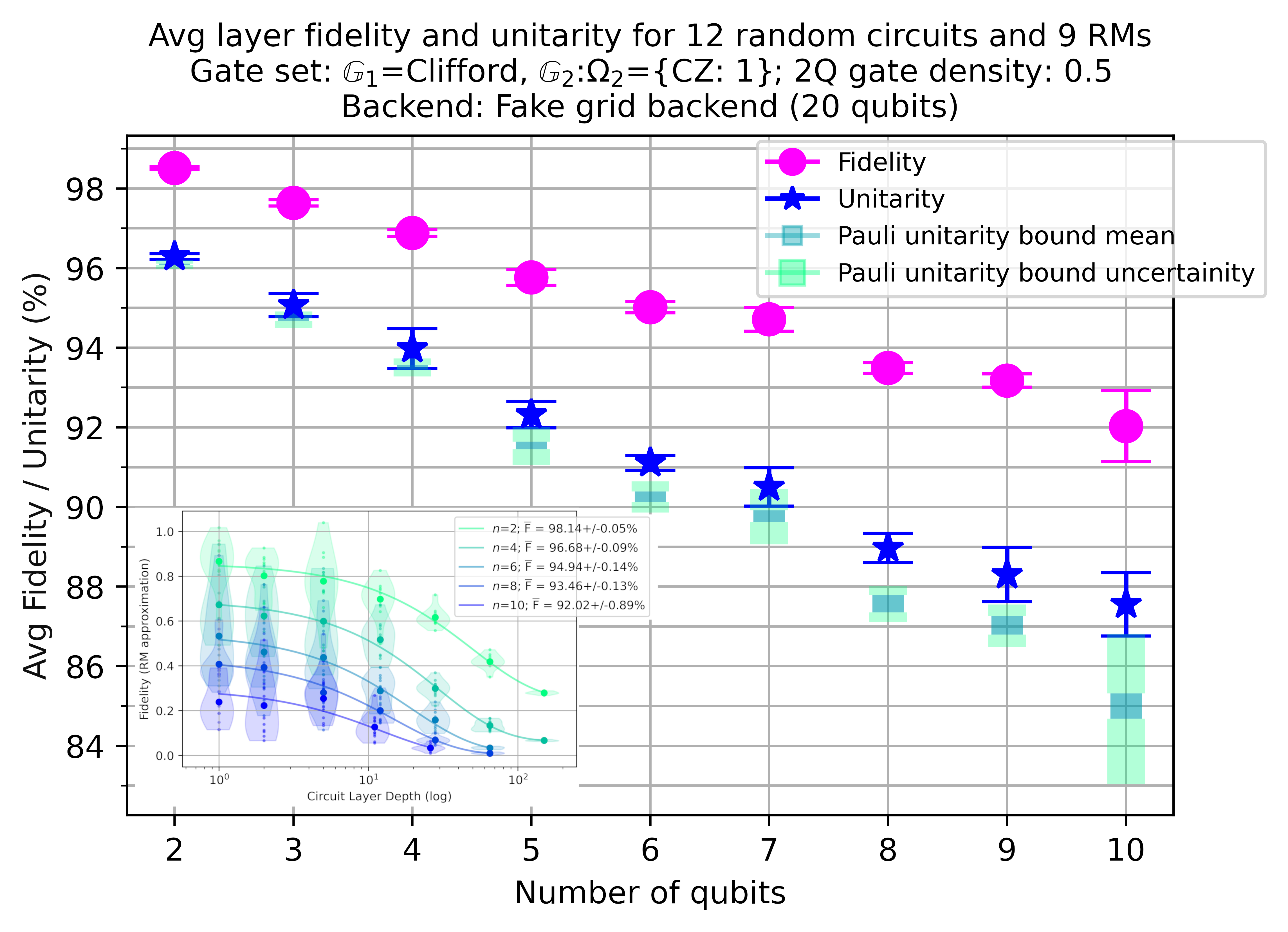}
    \caption{\textbf{Average layer fidelities and unitarities in number of qubits for simulated backend}. Qubit layouts correspond to those of the inset of Fig.~\ref{fig: purity} (for odd $n$, the smaller qubit label was taken); data corresponding to the same $\Omega$-distributed circuits with gate ensemble $\mbb{G}=\mathsf{Clif}_1\cup\{\mathsf{CZ}\}$. In the inset, we show corresponding fidelity decays for even $n$, where individual plots correspond to a given number of qubits, small points to \gls{moms} estimators of Eq.~\eqref{eq: main average fidelity estimation in depth} at given depths, larger points to the respective average purity, and lines to least-squares fit of the averages to the model in Eq.~\eqref{eq: fidelity decay}, whereby the corresponding average layer fidelity $\overline{\gatefid}$ for respective $\mathsf{n}$ is extracted according to Eq.~\eqref{eq: average polarization}. Sample parameters: $N_{\mathsf{C}_m^*}=12$, $N_\mathsf{W}=9$, $N_{\mathsf{meas}}=2^{10}$ and $K=9$ \gls{moms}. Pauli unitarity bounds computed according to Ineq.~\eqref{eq: main Pauli unitarity bounds}.}
    \label{fig: unitarity and fidelity}
\end{figure*}

 In Fig.~\ref{fig: unitarity and fidelity} we show all average layer fidelities and unitarities in increasing number of qubits, together with the Pauli noise unitarity intervals as predicted by the bound in Ineq.~\eqref{eq: main Pauli unitarity bounds}. In the corresponding inset, we show the average state fidelity decays in increasing depth for each number of qubits $n$, and thus the corresponding average gate fidelities; similar to the case of \spark, these were computed using the measurements within the same experiment and using the estimator in Eq.~\eqref{eq: main average fidelity estimation in depth} using the measurements of the noiseless circuits. The noiseless probabilities were estimated with the same number of $N_{\mathsf{meas}}=2^{10}$ shots per measurement.

Generally, while variances in the decays are relatively high, the averages show only a modest deviation up to 9 qubits. The noise model is dominated by stochastic errors, so it is expected that most points should fall within the Pauli unitarity bound; the cases of $n=6,8$ are indeed outliers, which in experiment would need to be investigated further, in our case they most likely can be explained by both the qubit arrangement and the amplitude damping $T_1$ parameters involved. It should be noted that for simulation in a common laptop, going beyond 10 qubits becomes quite computationally demanding.

\subsection{Comparison of layer fidelity estimations with MRB}
The $\Omega$-distributed circuits we consider are a primitive component in \gls{mrb}~\cite{hines2022demonstrating}, thus the average fidelity estimations from both our technique and \gls{mrb} should agree. There are two main differences between both procedures: one is that we do not construct a ``mirror'' circuit, i.e., we do not append the circuit made up of the inverses of the original $\Omega$-distributed circuit, and the second is that we do not interleave layers of random Pauli gates. Our technique is able to estimate average unitarity too, although this comes at a sampling complexity cost that requires exponentially many more samples to increase accuracy.

\subsubsection{\texorpdfstring{\spark}{IQM Spark (TM)}}\label{appendix: mrb comparison spark}
\begin{figure}[ht!]
    \centering
\includegraphics[width=\textwidth]{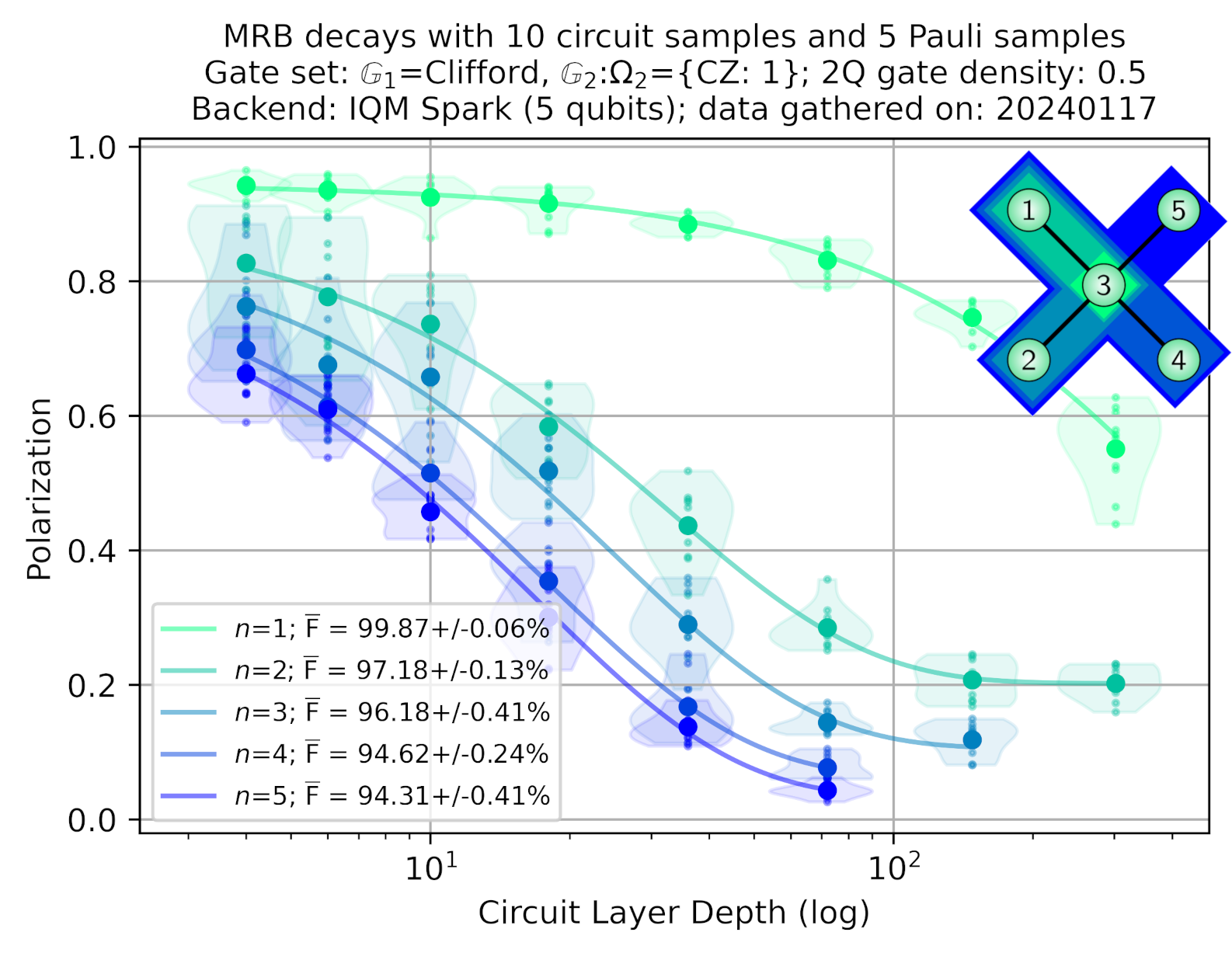}
    \caption{\textbf{Comparison with MRB on \spark}: For the same $\Omega$-distributed circuits considered in \S~\ref{sec: demonstration}}
    \label{fig: mrb spark}
\end{figure}

Tthe estimated fidelities from randomized measurements should be consistent with those estimated by \gls{mrb}~\cite{hines2022demonstrating}. Indeed, a primitive of \gls{mrb} is $\Omega$-distributed circuits; a difference with our technique is that we do not implement a mirror (i.e. the sequence of inverses), nor interleave random Pauli operators. Nevertheless, both techniques can estimate the same quantity (\gls{mrb} through polarizations instead of state fidelities), \gls{rm}s simply enable to access the unitarity. In Fig.~\ref{fig: mrb spark} we show the fidelity decays obtained experimentally with \gls{mrb} about a month ago on \spark, with the same $\Omega$-distributed circuits considered in \S~\ref{sec: demonstration}, namely made up of single-qubit Clifford gates uniformly sampled and $\mathsf{CZ}$ gates with a two-qubit gate density of $1/2$ according to the edge-grab sampler.

Clearly, while the estimated average layer fidelities agree, the distributions have a smaller variance than those in Fig.~\ref{fig: fidelity spark}, and were obtained with less total circuit samples. It can be conceivable then, that alternatively, average fidelities may be computed by constructing mirror circuits and following the \gls{mrb} protocol~\cite{hines2022demonstrating}, at the expense of performing separate measurements of these circuits.

\section{Unitarity and randomized compiling}\label{appendix: Pauli stuff}
To begin with, let us label the $n$-qubit Pauli group $\mbb{P}_n$, i.e., the group made up of all $n$-fold products of Pauli operators ($\mbb1,X,Y,Z$), as in~\cite{aces}: let $\boldsymbol{a}$ be a $2n$-bit string, $\boldsymbol{a}=a_1a_2\ldots{a}_{2n}$, and define
\begin{equation}
    P_{\boldsymbol{a}} := i^{\boldsymbol{a}^\mathrm{T}\Upsilon\boldsymbol{a}}\prod_{j=1}^nX_j^{a_{2j-1}}Z_j^{a_{2j}},
\end{equation}
where $X_j$ and $Z_j$ are single-qubit Pauli operators acting on the $j$\textsuperscript{th} qubit and $\Upsilon:=\Moplus_{j=1}^n(\begin{smallmatrix}0&1\\0&0\end{smallmatrix})$ such that $P_{\boldsymbol{a}}$ is ensured to be Hermitian. The Pauli group $\mbb{P}_n$ is then made up of all such $P_{\boldsymbol{a}}$ together with their products and their overall phases $\{\pm1,\pm{i}\}$; e.g., $\mbb{P}_1=\{\pm\mbb1,\pm{i}\mbb1,\pm{X},\ldots,\pm{i}Z\}$. In particular, denoting any other $2n$-bit strings with bold lowercase letters, we will make use of the property
\begin{equation}
    P_{\boldsymbol{a}}P_{\boldsymbol{b}} = (-1)^{\langle\boldsymbol{a},\boldsymbol{b}\,\rangle}P_{\boldsymbol{b}}P_{\boldsymbol{a}},
    \label{eq: p1 Paulis}
\end{equation}
where
\begin{align}
    \langle\boldsymbol{a},\boldsymbol{b}\,\rangle &:= \boldsymbol{a}^\mathrm{T}(\Upsilon+\Upsilon^\mathrm{T})\boldsymbol{b}\mod{2} \nonumber\\
    &= (a_1b_2 + a_2b_1 + a_3b_4 + a_4b_3 + \ldots+a_{2n}b_{2n-1}+a_{2n-1}b_{2n})\mod2,
\end{align}
which in particular is such that
\begin{equation}
    \f{1}{4^n}\sum_{\boldsymbol{a}}(-1)^{\langle\boldsymbol{a},\boldsymbol{b}+\boldsymbol{c}\,\rangle} = \delta_{\boldsymbol{b}\boldsymbol{c}}\,.
    \label{eq: p2 Paulis}
\end{equation}

Let us then write a generic $n$-qubit noise channel (i.e., some \gls{cp} map) in the so-called $\chi$-representation, $\mc{E}(\cdot)=\sum_{\boldsymbol{a},\boldsymbol{b}}\chi_{\boldsymbol{a}\boldsymbol{b}}P_{\boldsymbol{a}}(\cdot){P}_{\boldsymbol{b}}$, where $\chi$ is a positive Hermitian matrix. We can also relate this representation to the so-called \gls{ptm} representation, which is a $4^n$-square matrix with real entries given by
\begin{align}
    \boldsymbol{\mc{E}}_{\boldsymbol{ab}} &:= \f{1}{2^n}\tr[P_{\boldsymbol{a}}\mc{E}(P_{\boldsymbol{b}})] \nonumber\\
    &= \f{1}{2^n}\sum\chi_{\boldsymbol{{cd}}}\tr[P_{\boldsymbol{a}}{P}_{\boldsymbol{c}}P_{\boldsymbol{b}}P_{\boldsymbol{d}}],
\end{align}
which is not quite informative by itself; however, a particular case where this representation is useful is for the case of a Pauli channel, which has for $\chi$ a diagonal matrix, implying that its \gls{ptm} is diagonal too. A Pauli channel can be enforced on any channel by twirling it with the Pauli group, i.e.,
\begin{align}
    \mc{E}\,\mapsto\,\mc{E}^{\twist}(\cdot) := \f{1}{4^n}\sum_{P\in\mbb{P}_n}P\mc{E}(P\cdot{P})P,
    \label{eq: Pauli twirl exact}
\end{align}
which gives
\begin{align} \boldsymbol{\mc{E}}^{\twist}_{\boldsymbol{ab}} &= \f{1}{2^{3n}} \sum_{\boldsymbol{c,d,q}} \chi_{\boldsymbol{{cd}}}\tr[P_{\boldsymbol{a}}P_{\boldsymbol{q}}{P}_{\boldsymbol{c}}P_{\boldsymbol{q}}P_{\boldsymbol{b}}P_{\boldsymbol{q}}P_{\boldsymbol{d}}P_{\boldsymbol{q}}] \nonumber\\
    &= \f{1}{2^{3n}} \sum_{\boldsymbol{c,d,q}}   (-1)^{\langle\boldsymbol{q},\boldsymbol{c}+\boldsymbol{d}\rangle}\,\chi_{\boldsymbol{{cd}}}\,\tr[P_{\boldsymbol{a}}{P}_{\boldsymbol{c}}P_{\boldsymbol{b}}P_{\boldsymbol{d}}] \nonumber\\
    &= \f{1}{2^n} \sum_{\boldsymbol{c,d}}  \chi_{\boldsymbol{{cd}}} \delta_{\boldsymbol{c}\boldsymbol{d}}\tr[P_{\boldsymbol{a}}{P}_{\boldsymbol{c}}P_{\boldsymbol{b}}P_{\boldsymbol{d}}] \nonumber\\
    &= \f{1}{2^n} \sum_{\boldsymbol{d}} \chi_{\boldsymbol{{dd}}} \tr[P_{\boldsymbol{a}}{P}_{\boldsymbol{d}}P_{\boldsymbol{b}}P_{\boldsymbol{d}}] \nonumber\\
    &= \delta_{\boldsymbol{a}\boldsymbol{b}} \sum_{\boldsymbol{d}}  (-1)^{\langle\boldsymbol{a},\boldsymbol{d}\rangle}\chi_{\boldsymbol{{dd}}} := \delta_{\boldsymbol{a}\boldsymbol{b}} \sum_{\boldsymbol{d}}  (-1)^{\langle\boldsymbol{a},\boldsymbol{d}\rangle}\alpha_{\boldsymbol{{d}}},
\end{align}
where we made use of both Eq.~\eqref{eq: p1 Paulis} and Eq.~\eqref{eq: p2 Paulis}, so that indeed $\boldsymbol{\mc{E}}^{\twist}$, the \gls{ptm} of the twirled map $\mc{E}^{\twist}$, is manifestly diagonal with eigenvalues $\lambda_{\boldsymbol{a}}:=\sum_{\boldsymbol{b}}  (-1)^{\langle\boldsymbol{a},\boldsymbol{b}\rangle}\alpha_{\boldsymbol{b}}$, where we write the diagonal elements with a single index as $\alpha_{\boldsymbol{a}}:=\chi_{\boldsymbol{aa}}$.

The diagonal elements $\alpha$ are the so-called Pauli error probability rates, conversely related to the \gls{ptm} eigenvalues as $\alpha_{\boldsymbol{a}}=2^{-n}\sum_{\boldsymbol{b}}(-1)^{\langle\boldsymbol{a},\boldsymbol{b}\rangle}\lambda_{\boldsymbol{b}}$, and in terms of these, the trace non-increasing condition translates to $\sum_{\boldsymbol{a}}\alpha_{\boldsymbol{a}}\leq1$, saturating for $\mc{E}$ being \gls{tp}.

If follows that the average gate fidelity of a Pauli channel is given by
\begin{align}
    \overline{\gatefid}(\mc{E}^{\twist}) &= \f{\tr[\boldsymbol{\mc{E}}^{\twist}]+2^n}{2^n(2^n+1)} \nonumber\\
    &= \f{\sum_{\boldsymbol{a}}\lambda_{\boldsymbol{a}}+2^n}{2^n(2^n+1)} \nonumber\\
    &= \f{2^n\alpha_{\boldsymbol{0}}+1}{2^n+1}\nonumber\\
    &= \overline{\gatefid}(\mc{E}) \nonumber\\
    &\stackrel{\footnotesize{\mc{E}\text{ is TP}}}{=} 1 - \left(\f{2^n}{2^n+1}\right)\sum_{\boldsymbol{i}\neq\boldsymbol0}\alpha_{\boldsymbol{i}},
\end{align}
where in the penultimate line we highlighted that it is the same as for the raw (not-twirled) channel $\mc{E}$, and where used Eq.~\eqref{eq: p2 Paulis}, with $\boldsymbol{0}$ being the $2n$-bit string with all bits being zero, and in the last line the case $\sum_{\boldsymbol{a}}\alpha_{\boldsymbol{a}}=1$ for $\mc{E}$ being \gls{tp}; such expression makes it manifest that average gate fidelity captures only the information of the probability of \emph{any} Pauli error happening~\footnote{ This does \emph{not} mean that average gate fidelity is insensible to coherent errors, but rather that it is only partially so through their contribution to the diagonal elements in the \gls{ptm}.}.

Conversely, the unitarity of the Pauli channel is given by
\begin{align}
    \overline{\unitarity}(\mc{E}^{\twist}) &= \f{\tr[\boldsymbol{\mc{E}}^{\twist\,\dg}\boldsymbol{\mc{E}}^{\twist}]-1}{4^n-1} \nonumber\\
    &= \f{\sum_{\boldsymbol{a}}\lambda_{\boldsymbol{a}}^2-1}{4^n-1} \nonumber\\
    &= \f{4^n\sum_{\boldsymbol{a}}\alpha_{\boldsymbol{a}}^2-1}{4^n-1},
\end{align}
which is minimal in the sense that for a non-identity it corresponds to a purely stochastic noise channel, as it only involves the diagonal $\chi$-matrix terms, the Pauli error rates, involved in the average gate-fidelity. Clearly, when the distribution of errors is uniform, i.e., all $\alpha_{\boldsymbol{i}}=2^{-n}$ so that the channel is maximally depolarizing, the unitarity vanishes. As opposed to average gate fidelity, the average unitarity of the raw channel $\mc{E}$ would capture all elements of its \gls{ptm}. We can further lower-bound this as follows,
\begin{align}
    \overline{\unitarity}(\mc{E}^{\twist}) &= \f{4^n\sum_{\boldsymbol{a}}\alpha_{\boldsymbol{a}}^2-1}{4^n-1} \nonumber\\
    &\leq \f{4^n\left[\alpha_{\boldsymbol{0}}^2+\left(\sum_{\boldsymbol{b}\neq\boldsymbol{0}}\alpha_{\boldsymbol{b}}\right)^2\right]-1}{4^n-1} \nonumber\\
    &\leq \f{4^n\left[\alpha_{\boldsymbol{0}}^2+(1-\alpha_{\boldsymbol{0}})^2\right]-1}{4^n-1} \nonumber\\
    &= \left(\f{2^n\overline{\gatefid}(\mc{E})-1}{2^n-1}\right)^2 + (4^n-2)\left(\f{1-\overline{\gatefid}(\mc{E})}{2^n-1}\right)^2,
\end{align}
so together with the lower bound $\overline{\unitarity}\overline{f}(\mc{E})^2$ of~\cite{Wallman_2015}, we can express the unitarity for the Pauli-twirled channel $\mc{E}^{\twist}$ as
\begin{equation}
    \overline{f}(\mc{E})^2 \leq \overline{\unitarity}(\mc{E}^{\twist}) \leq \overline{f}(\mc{E})^2 + \f{2^{2n}-2}{(2^n-1)^2}\,\overline{r}(\mc{E})^2,
    \label{eq: upper bound on unitarity}
\end{equation}
where here $\overline{f}(\mc{E})=(2^n\overline{\gatefid}(\mc{E})-1)/(2^n-1)$ is the noise-strength parameter in Eq.~\eqref{eq: average polarization} in the main text, and $\overline{r}(\mc{E}):=1-\overline{\gatefid}(\mc{E})$ is the average gate infidelity of $\mc{E}$.

The lower bound in Ineq.~\eqref{eq: upper bound on unitarity} saturates for depolarizing noise, i.e., when all non-identity Pauli error rates are the same, while the upper bound overestimates the unitarity for Pauli noise through cross-products of non-identity Pauli error rates. Thus, the upper bound serves as a proxy for whether noise has coherent components, if it is above it, or whether it is potentially Pauli, if it is under it.

Of course, in practice, we cannot directly twirl noise quantum channels but we need a way of achieving it by manipulating the noisy gates in question. Whenever the ideal (or target) gates are Clifford, a Pauli twirl can be enforced by a so-called $\mc{G}$-twisted twirl~\cite{aces}, where the outer Pauli operators in Eq.~\eqref{eq: Pauli twirl exact} are acted on (or twisted) with the ideal Cliffords. This concept of a twisted-twirl, together with employing random samples instead of all the $4^n$ possible Pauli operators, are at the core of so-called \gls{rc}, where we would then effectively have
\begin{align}
    \mc{E}^{\twist}_N(\cdot) := \f{1}{N}\sum_{P\sim\mbb{P}_n}^{N\,\text{samples}}P\mc{E}(P\cdot{P})P,
\end{align}
where here $P\sim\mathsf{Pauli}$ means $P$ are Pauli samples drawn uniformly at random with no repetition from $\mbb{P}_n$, and the sum runs over $N\leq4^n$ such samples. In practice, this randomized sum would be computed by measuring and then averaging many circuits where the noisy gate is twisted-twirled; considering that the noisy gate may be transpiled into single and two-qubit gates, the way this is usually done is by precisely compiling the single-qubit gates with the Paulis in each random sample, with the overall gate being logically exactly the same. Since also in practice the Pauli operators will be noisy, care has to be taken not to increase the overall number of gates that are executed. In general if the Paulis themselves have coherent noise, perfect diagonalization would be achieved strictly for $N\to\infty$ by sampling with repetition.

The corresponding \gls{ptm} of $\mc{E}^{\twist}_N(\cdot)$ has components
\begin{align}
    \left(\boldsymbol{\mc{E}}^{\twist}_N\right)_{\boldsymbol{ab}} &= \f{1}{N2^{n}} \sum_{\boldsymbol{q}}^{N\,\text{samples}}\sum_{\boldsymbol{c,d}} \chi_{\boldsymbol{{cd}}}\tr[P_{\boldsymbol{a}}P_{\boldsymbol{q}}{P}_{\boldsymbol{c}}P_{\boldsymbol{q}}P_{\boldsymbol{b}}P_{\boldsymbol{q}}P_{\boldsymbol{d}}P_{\boldsymbol{q}}] \nonumber\\
    &= \f{1}{2^{n}} \sum_{\boldsymbol{c,d}}   \left(\f{1}{N}\sum_{\boldsymbol{q}}^{N\,\text{samples}}(-1)^{\langle\boldsymbol{q},\boldsymbol{c}+\boldsymbol{d}\rangle}\right)\,\chi_{\boldsymbol{{cd}}}\,\tr[P_{\boldsymbol{a}}{P}_{\boldsymbol{c}}P_{\boldsymbol{b}}P_{\boldsymbol{d}}],
\end{align}
where now we can only approximate the property in Eq.~\eqref{eq: p2 Paulis} with the $N$ random samples of bit strings $\boldsymbol{q}$: in the infinite sample limit, $N\to\infty$, ,or if exactly each distinct bit string is sampled once, the term within parenthesis goes to $\delta_{\boldsymbol{cd}}$. Since $\langle\boldsymbol{q},2\boldsymbol{c}\rangle=0$ for any pair of bit strings $\boldsymbol{q}$ and $\boldsymbol{c}$, the sum over $\boldsymbol{q}$ when $\boldsymbol{c}=\boldsymbol{d}$ equals 1 for any number of samples; this just reflects the fact that twirling (or \gls{rc}) leaves the diagonal of the \gls{ptm} invariant. Thus,  we can write
\begin{align}
\left(\boldsymbol{\mc{E}}^{\twist}_N\right)_{\boldsymbol{ab}} &= \boldsymbol{\mc{E}}^{\twist}_{\boldsymbol{ab}} + \sum_{\boldsymbol{c}\neq\boldsymbol{d}} \left(\f{1}{N}\sum_{\boldsymbol{q}}^{N\,\text{samples}}(-1)^{\langle\boldsymbol{q},\boldsymbol{c}+\boldsymbol{d}\rangle}\,\chi_{\boldsymbol{{cd}}}\,\f{\tr[P_{\boldsymbol{a}}{P}_{\boldsymbol{c}}P_{\boldsymbol{b}}P_{\boldsymbol{d}}]}{2^n}\right) \nonumber\\
&:= \boldsymbol{\mc{E}}^{\twist}_{\boldsymbol{ab}} + \f{1}{N}\sum_{\boldsymbol{q}}^{N\,\text{samples}}\tilde{\lambda}_{\boldsymbol{ab}}^{(\boldsymbol{q})},
\end{align}
where here $\mc{E}^{\twist}$ as before, is the perfectly Pauli-twirled channel in Eq.~\eqref{eq: Pauli twirl exact}, and where we defined 
\begin{equation}
\tilde{\lambda}_{\boldsymbol{ab}}^{(\boldsymbol{q})}:=\sum_{\boldsymbol{c}\neq\boldsymbol{d}}(-1)^{\langle\boldsymbol{q},\boldsymbol{c}+\boldsymbol{d}\rangle}\,\chi_{\boldsymbol{{cd}}}\,\f{\tr[P_{\boldsymbol{a}}{P}_{\boldsymbol{c}}P_{\boldsymbol{b}}P_{\boldsymbol{d}}]}{2^n},
\end{equation}
which correspond to the off-diagonal elements of the \gls{ptm} upon the twirl action of an element $P_{\boldsymbol{q}}$. The effect of $N=1$, a single randomization, will simply be to either change a sign of an off-diagonal element or leave it as it is. For an even number $N$ of samples, the off-diagonals will either vanish or be suppressed in magnitude as integer multiples of $2/N$ (with ratio less than one). On the other hand, for odd $N$, all off-diagonals vanish in magnitude as integer multiples of $1/N$.

Finally, then
\begin{align}
    \overline{\unitarity}(\mc{E}^{\twist}_N) &= \overline{\unitarity}(\mc{E}^{\twist}) + \f{1}{N^2}\sum_{\boldsymbol{q},\boldsymbol{q}^\prime}^{N\,\text{samples}}\f{\tilde{\lambda}_{\boldsymbol{ab}}^{(\boldsymbol{q})}\tilde{\lambda}_{\boldsymbol{ba}}^{(\boldsymbol{q}^\prime)}}{4^n-1},
\end{align}
where similarly now, the off-diagonal contributions will vanish as (small) multiples of $1/N^2$.

\end{document}